\documentclass{aa}
\usepackage{epsfig}

\newbox\grsign \setbox\grsign=\hbox{$>$} \newdimen\grdimen \grdimen=\ht\grsign
\newbox\simlessbox \newbox\simgreatbox \newbox\simpropbox \newbox\wtildebox 
\setbox\simgreatbox=\hbox{\raise.5ex\hbox{$>$}\llap
     {\lower.5ex\hbox{$\sim$}}}\ht1=\grdimen\dp1=0pt
\setbox\simlessbox=\hbox{\raise.5ex\hbox{$<$}\llap
     {\lower.5ex\hbox{$\sim$}}}\ht2=\grdimen\dp2=0pt
\def\simgreat{\mathrel{\copy\simgreatbox}}

\newcommand{\Msun}{\mbox{$M_{\odot}$}}

\newcommand{\Msunpc}{\mbox{$M_{\odot}$~pc$^{-3}$}}

\newcommand{\Mcl}{\mbox{$M_{\rm cl}$}}
\newcommand{\be}{\mbox{\begin{equation}}}
\newcommand{\ee}{\mbox{\end{equation}}}

\newcommand{\Mmin}{\mbox{$M_{\rm min}$}}
\newcommand{\Mmax}{\mbox{$M_{\rm max}$}}

\newcommand{\Mi}{\mbox{$M_i$}}
\newcommand{\Mp}{\mbox{$M$}}

\newcommand{\Mpt}{\mbox{$M(t;M_{\rm i})$}}

\newcommand{\tdis}{\mbox{$t_{\rm dis}$}}
\newcommand{\tdisBM}{\mbox{$t_{\rm dis}^{\rm BM}$}}

\newcommand{\Cref}{\mbox{$m_{\rm ref}$}}

\newcommand{\cfr}{\mbox{$CFR(t)$}}
\newcommand{\nmit}{\mbox{$N(M_i,t)$}}

\newcommand{\qev}{\mbox{$q_{\rm ev}$}}
\newcommand{\qevt}{\mbox{$q_{\rm ev}(t)$}}
\newcommand{\mut}{\mbox{$\mu(t;M_{\rm i})$}}
\newcommand{\Mptmax}{\mbox{$M_{\rm up}(t)$}}
\newcommand{\tupmp}{\mbox{$t_{\rm up}(M)$}}
\newcommand{\tup}{\mbox{$t_{\rm up}$}}
\newcommand{\tgal}{\mbox{$t_0$}}

\newcommand{\ttot}{\mbox{$t_{\rm dis}^{\rm total}$}}
\newcommand{\tninefive}{\mbox{$t^{\rm 95\%}_{\rm dis}$}}
\newcommand{\tdelta}{\mbox{$t_{\rm dis}^{1-\Delta}$}}

\newcommand{\muevol}{\mbox{$\mu_{\rm ev}$}}
\newcommand{\muevt}{\mbox{$\mu_{\rm ev}(t)$}}

\newcommand{\nbody}{\mbox{$N$-body}}
\newcommand{\Nbody}{\mbox{$N$-body}}

\begin{document}

\title{An analytical description of the disruption of star clusters
in tidal fields with an application to Galactic open clusters}

\author{Henny J.G.L.M. Lamers \inst{1,2}, Mark Gieles\inst{1},
        Nate Bastian \inst{1},
        Holger Baumgardt\inst{3},        
        Nina V. Kharchenko\inst{4,5,6}
        and Simon Portegies Zwart\inst{7,8}
        }

\institute { 
                 {Astronomical Institute, Utrecht University, 
                 Princetonplein 5, NL-3584CC Utrecht, the Netherlands
                 {\tt  lamers@astro.uu.nl, bastian@astro.uu.nl,  
                 gieles@astro.uu.nl}}
            \and  {SRON Laboratory for Space Research, Sorbonnelaan 2,
                 NL-3584CC, Utrecht, the Netherlands}
            \and {Sternwarte, University of Bonn, Auf dem H\"ugel 71,
                  DE-53121 Bonn, Germany 
                  {\tt holger@astro.uni-bonn.de}}
            \and {Astrophysikalisches Institut Potsdam, An der Sternwarte 16,
                  D-14482 Potsdam, Germany
                  {\tt nkharchenko@aip.de}}
             \and {Astronomisches Rechen-Institut, M\"onchhofstra{\ss}e 12-14,
                  D-69120 Heidelberg, Germany
                  {\tt nkhar@ari.uni-heidelberg.de}}
             \and {Main Astronomical Observatory, 27 Academica Zabolotnogo Str.,
                  03680  Kiev, Ukraine
                  {\tt nkhar@mao.kiev.ua}}
            \and {Astronomical Institute, University of Amsterdam,
                 Kruislaan 403, NL-1098SJ, Amsterdam, the Netherlands
                 {\tt spz@science.uva.nl}}
            \and {Informatics Institute, University of Amsterdam,
                 Kruislaan 403, NL-1098SJ, Amsterdam, the Netherlands
                 }
            }

\date{Received date ; accepted date}

\offprints{H. J. G. L. M. Lamers}

\abstract{
We present a simple analytical description of the disruption of star
clusters in a tidal field. 
The cluster disruption time, defined as $\tdis = \{d \ln ~M/dt\}^{-1}$, 
depends on the mass $M$ of the cluster as 
$\tdis =\tgal (M/\Msun)^{\gamma}$ with $\gamma=0.62$ for clusters in a
tidal field, as shown by 
empirical studies of cluster samples in different galaxies
and by \nbody\ simulations.
Using this simple description  we derive an analytic expression
for the way in which the mass of a cluster decreases with time due to
stellar evolution and disruption.
The result agrees excellently with those of detailed \nbody\
simulations for clusters in the tidal field of our galaxy.
The analytic expression can be used to predict the mass and age
histograms of surviving clusters for any cluster initial mass function
and any cluster formation history. 
The method is applied to explain the age distribution of the
open clusters in the solar neighbourhood within 600 pc, based on the new
cluster sample of Kharchenko et al. that appears to be unbiased
within a distance of about 1 kpc. From a comparison between the
observed and predicted age distributions in the age range between 10 Myr to
3 Gyr we find the following results:
(1) The disruption time of a $10^4$ \Msun\ cluster in the solar
neighbourhood is about $1.3 \pm 0.5$ Gyr.
This is a factor 5 shorter than derived from $N$-body simulations
of clusters in the tidal field of the galaxy. Possible reasons for
this discrepancy are discussed.
(2) The present starformation rate in bound clusters 
within 600 pc from the Sun is $5.9 \pm 0.8~10^2$ \Msun Myr$^{-1}$,
which corresponds to a surface star formation rate of bound clusters 
$5.2 \pm 0.7 \times 10^{-10}$ \Msun~yr$^{-1}$pc$^{-2}$. 
(3) The age distribution of open clusters shows a bump between 
0.26 and 0.6 Gyr when the cluster formation rate was 2.5 times higher
than before and after. 
(4) The present star formation rate in bound
clusters is about half as small as that derived from the 
study of embedded clusters.
The difference suggests that about half of the clusters in the solar
neighbourhood become unbound within about 10 Myr.
(5) The most massive clusters within 600 pc had an initial
mass of about $3~10^4$ \Msun. This is in agreement with
the statistically expected value based on a cluster initial mass function
with a slope of -2, even if the physical upper mass limit for cluster
formation is as high as $10^6$ \Msun.
}

\authorrunning{H.J.G.L.M. Lamers et al.}
\titlerunning{Disruption of star clusters in tidal fields} 

\maketitle


\section{Introduction}
\label{sec:1}

Bound star clusters\footnote{Bound clusters are those that survive 
the infant mortality due to the removal of gas during the first $10^7$
years (Fall 2004; Bastian et al. 2005)}
  in a tidal field are losing mass due to 
internal effects, i.e. mass loss by stellar evolution, and by the external
effect of tidal stripping. The combination of these effects results 
in a decreasing mass of the cluster until the cluster is destroyed
completely. The time scale of this disruption depends on the initial
conditions of the cluster, e.g. the stellar initial mass function and
its concentration, and on the tidal forces experienced by the cluster
during its galactic orbit. 

The disruption of star clusters determines the mass and age
distributions of the existing clusters. Therefore any study of the 
cluster formation history of a galaxy has to take into account the 
disruption of clusters. 

The age and mass distribution of the
{\it initial} cluster population is described 
by the cluster formation rate as a function of time, CFR(t), and the 
cluster initial mass function, CIMF. The distributions 
of the {\it present observable} star clusters is modified
because: (a) the disruption time of the clusters depends on the
initial mass, (b) the mass of each cluster decreases with time,
(c) the clusters fade as they age due to stellar evolution.
A (simple) description of these three effects would facilitate the
studies of samples of star clusters. The purpose of this paper is to
provide such a simple description and show how it can be used in the
analysis of star cluster samples. 

The structure of the paper is as follows:\\
In Sect. 2 we discuss the arguments that the cluster disruption time 
depends on its mass as a power law of the type $\tdis \propto M^{0.62}$.
In Sect. 3 we describe the expressions used to calculate the evolution of a 
cluster in terms of its decreasing mass due to stellar evolution and disruption.
In Sect. 4 we will show that the results agree very well with those
of \nbody\ simulations of clusters in a tidal field.
In Sect. 5 we use the description of the decreasing mass to predict
the mass and age distributions of cluster samples with
various initial mass functions and various cluster formation histories.
In Sect. 6 we apply the method by comparing the predicted age distribution
of open clusters in the solar neighbourhood to the observed sample
from Kharchenko et al. (2005). From this comparison we derive the
disruption time of open clusters in the solar neighbourhood as well
as the cluster formation rate and the star formation rate. We compare the
results with independent determinations.
The discussion and conclusions are in Sect. 7.

\section{A power law expression for the disruption time of star
  clusters}

Boutloukos \& Lamers (2003, hereafter BL03) have studied the mass and age
distributions of magnitude limited cluster samples in selected
regions in four galaxies, and concluded that these distributions
can be explained if the disruption time of clusters depends on the initial
 mass \Mi\ of the clusters as $\Mi^{\gamma}$,  with
$\gamma \simeq 0.6$ for clusters in very different local environments.

Baumgardt \& Makino (2003, hereafter BM03) have calculated a 
grid of \nbody\ simulations of clusters 
in the tidal field of our galaxy for different initial masses
and initial concentration factors in circular and elliptical
orbits at various galactocentric distances. They take into account
mass loss by stellar evolution and by tidal relaxation. Their
calculations show that the disruption time of a cluster, defined as
the time when 5\% of the initial number of stars remain in the
cluster, scales with the 
half mass relaxation time $t_{\rm rh}$ and the clusters crossing time
$t_{\rm cr}$ as $\tdis \propto t_{\rm rh}^x t_{\rm cr}^{1-x}$
with $x=0.82$ for clusters with an initial dimensionless depth
$W_0=7$ (which is a measure of the concentration index of the cluster,
see King 1966)
and $x=0.75$ for less concentrated clusters with $W_0=5$. 
BM03 and Gieles et al. (2004) have shown that for all the
models of BM03 the disruption time can be expressed as
a function of the initial cluster mass as

\begin{equation}
\tdis = \tgal~ (\Mi/\Msun)^{0.62}
\label{eq:tdis}
\end{equation}
where $\tgal$ is a constant that depends on the tidal field 
of the particular galaxy in which the
cluster moves and on the ellipticity of its orbit. 
So the predicted dependence of the disruption time on the initial mass
of a cluster agrees very well with the empirical relation derived by 
BL03. (De la Fuente Marcos \& de la Fuente Marcos (2004) also report
a power law dependence of the characteristic life time $\tau$ 
of clusters on the
number of stars. Based on a series of dynamical models they find that
$\tau \sim N^{0.68}$ where N is the initial number of stars of a cluster.) 
Lamers, Gieles \& Portegies Zwart (2005) showed that
$\tgal$ is expected to depend on the ambient density at the location of
the clusters in that galaxy as $\tgal \propto \rho_{\rm amb}^{-1/2}$.

The discussion above has concentrated on the comparison between the
disruption time of clusters of different initial  masses, i.e. $\tdis \propto
\Mi^{0.62}$,  within one galactic environment. We have not yet
discussed how the mass of an individual cluster decreases with time. This
is the topic of the next section.

\section{The decrease of the cluster mass due to stellar evolution and
tidal effects}

The mass of a cluster decreases due to stellar evolution and 
tidal disruption. We will describe the evolution of the bound mass, using
analytic expressions for the mass loss from the cluster by stellar evolution
and by tidal effects. 

\subsection{Mass loss by stellar evolution}

The mass loss from clusters due to stellar evolution has been
calculated for cluster evolution models by several groups,
e.g. Bruzual \& Charlot (1993) and the Starburst99 models by
Leitherer et al. (1999). We adopt the 
$GALEV$ models for single stellar populations (Schulz et al. 2002;
Anders \& Fritze-v. Alvensleben 2003). 
These models contain stars in the mass range of
$0.15 < M_* < 85$ \Msun, distributed over this mass range with either 
the Salpeter  or Scalo   mass function. 
We adopt the models with the Salpeter mass
function because deep photometry of clusters in the LMC shows that the
cluster IMF is a powerlaw with a slope of about -2.35 down to at least
0.6 \Msun\ (de Marchi, 2003). Lower mass stars hardly contribute to
the luminosity at ages less than 10 Gyr, but may contribute
significantly to the cluster mass. The $GALEV$ models are based
on stellar evolution tracks from the Padova group, which include mass
loss and overshooting (Bertelli et al. 1994; Girardi et al. 2000). 
Lamers (2005) has shown that the fraction of the initial
cluster mass, $M_i$, that is lost by stellar evolution in the $GALEV$ models, 
i.e $\qev \equiv (\Delta M)_{\rm ev}/\Mi$ where $(\Delta M)_{\rm ev}$
is the mass lost by stellar evolution, 
can be approximated very accurately by a function of the form

\begin{equation}
\log \qev(t)= (\log t-a_{\rm ev})^{b_{\rm ev}}+c_{\rm
  ev}~~{\rm for}~~ t>12.5~{\rm Myr} 
\label{eq:qevgot}
\end{equation}
The values of $a_{\rm ev}$, $b_{\rm ev}$ and $c_{\rm ev}$ are listed in Table
\ref{tbl:muevol} for different metallicities. 
This function describes the mass loss fraction of the models at
$t>12.5$ Myr with an accuracy of a few percent. The mass loss at 
younger ages is negligible because the most massive stars with $M_*>30
\Msun$ hardly contribute to the mass of the cluster.
For cluster models with a lower limit of the stellar IMF different
from 0.15 \Msun,  the
value of \qevt\ can easily be adjusted, because stars with $M<0.6 \Msun$
contribute to the cluster mass but not to its mass loss at ages less
than $10^{10}$ yrs.
(This mass loss rate is very different from that of the Starburst99
models, because the Starburst99 models have a lower limit for stellar 
mass of 1 \Msun.)

In this paper we use the symbol $\mu(t) \equiv M(t)/\Mi$ to describe the  
fraction of the mass of a cluster with initial mass \Mi\ that is still
bound at age $t$.
We define 
\begin{equation}
\muevt=1-\qevt
\label{eq:mevt}
\end{equation}
as the fraction of the initial mass 
of the cluster that would have remained at age  $t$,
if stellar evolution would have been the only mass loss mechanism.
The function $\muevt$ is independent of the initial mass of the cluster.

\begin{table}
\caption[]{Approximations to the mass lost by stellar evolution\\
for $GALEV$ cluster models with a Salpeter IMF of $\alpha=-2.35$, 
$0.15 < M_* < 85~\Msun$ and $0.0004<Z<0.05$}
\begin{tabular}{llll}
\hline\\
Z  & $a_{\rm ev}$ & $b_{\rm ev}$  & $c_{\rm ev}$\\
   &               &              & \\
\hline \\
0.0004 & 7.06 & 0.265 & -1.790   \\
0.0040 & 7.06 & 0.260 & -1.800   \\
0.0080 & 7.03 & 0.260 & -1.800   \\
0.0200 & 7.00 & 0.255 & -1.805   \\
0.0500 & 7.00 & 0.250 & -1.820   \\
\hline
\end{tabular}
\label{tbl:muevol}
\end{table}


\subsection{Mass loss by stellar evolution and tidal effects}

\begin{figure}[!ht]
\centerline{\psfig{figure=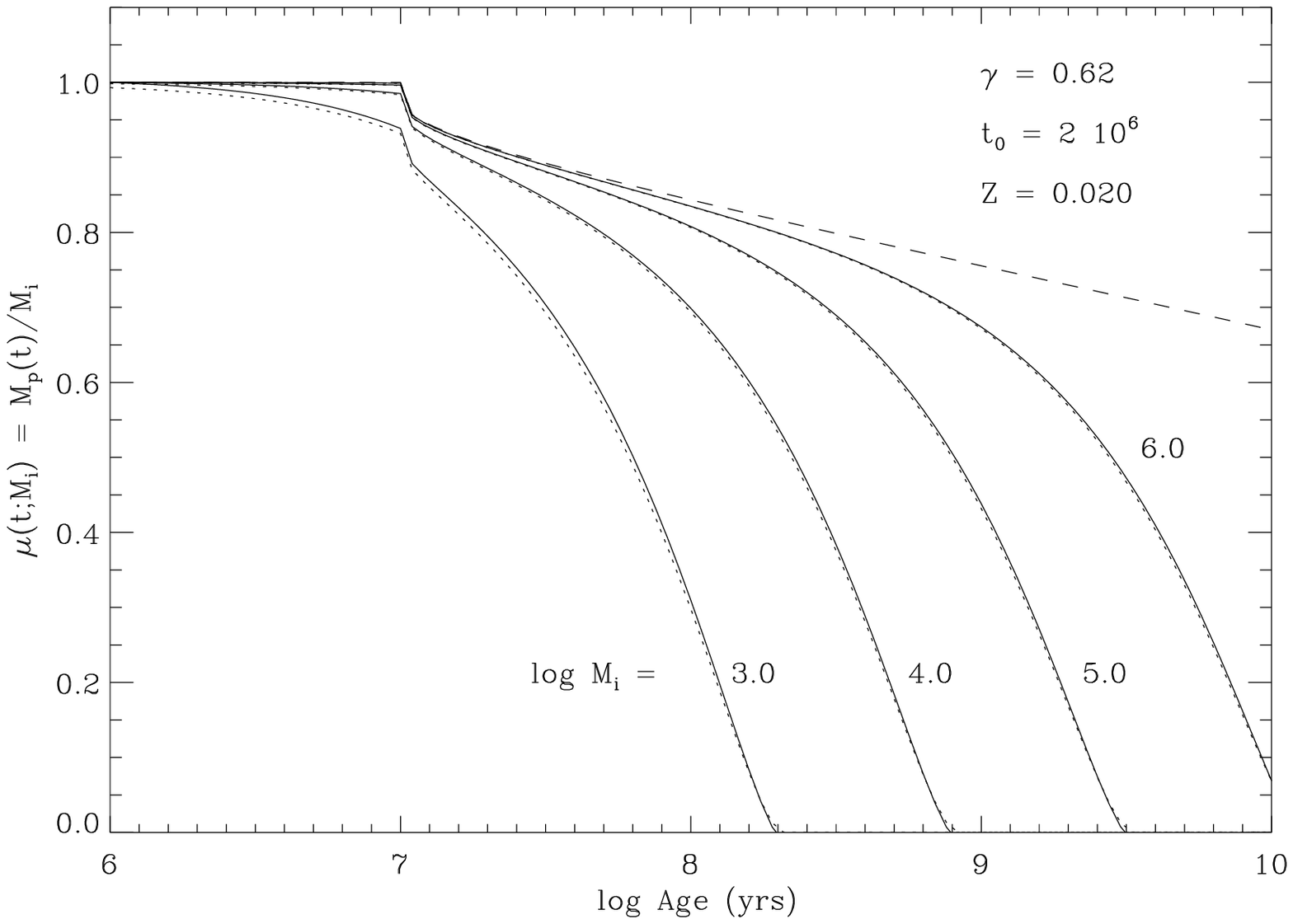,width=8.5cm}}
\centerline{\psfig{figure=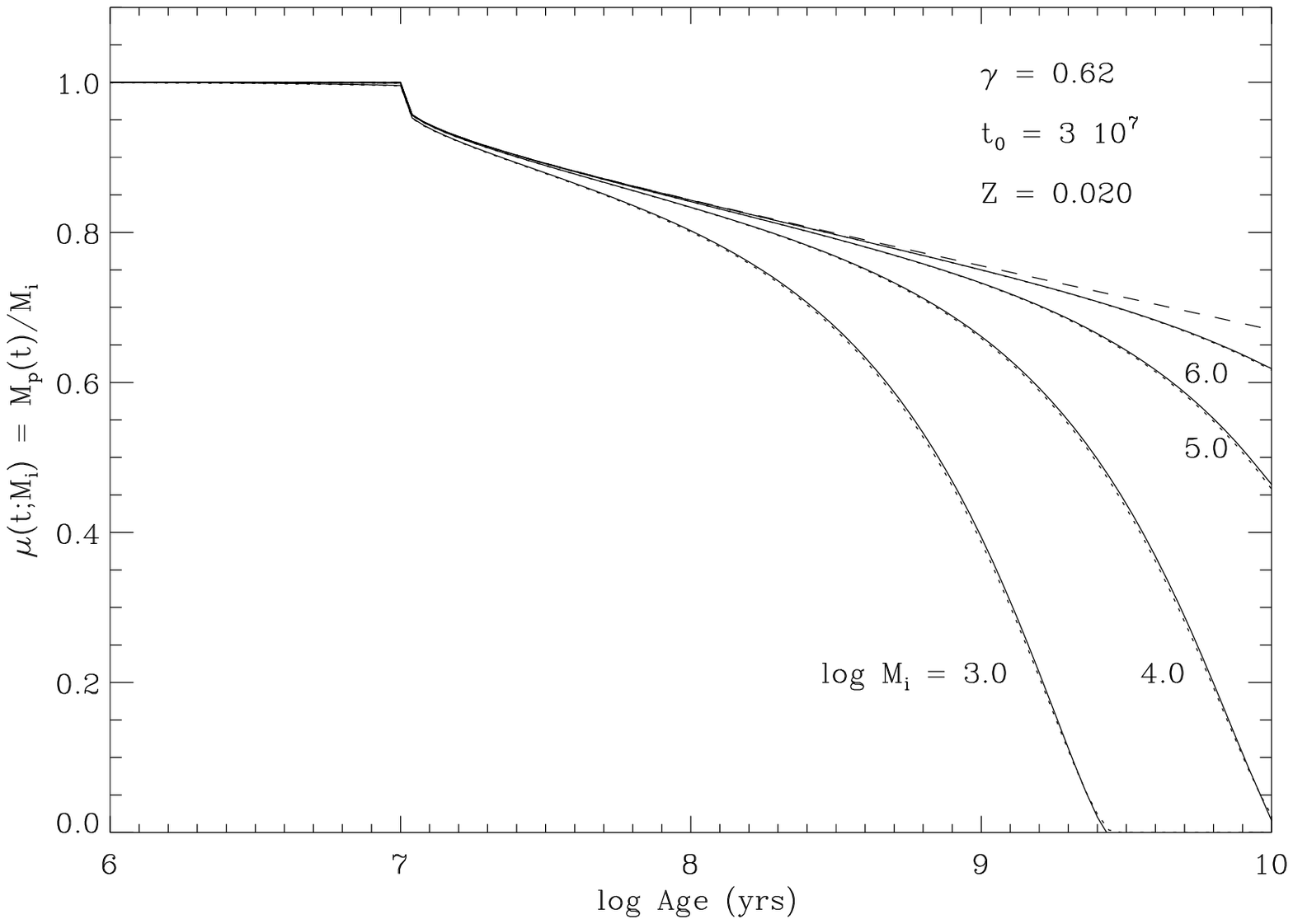,width=8.5cm}}

\caption[]{The predicted decrease in cluster mass due to stellar evolution and
  disruption for $Z=0.020$ and four values of the initial cluster
  masses: $10^3$, $10^4$, $10^5$ and $10^6~\Msun$.
  Top panel: $\tgal=2$ Myr;  lower panel: $\tgal=30$ Myr. 
  The full lines give the exact decrease derived by numerical
  solution of the differential equation (\protect{\ref{eq:dmpdt}}), 
  and the dotted lines give the approximation
  (Eq. \protect{\ref{eq:muapprox}}). Notice the excellent agreement.
  The dashed lines gives the decrease  in mass due to stellar evolution only. 
  }
\label{fig:1}
\end{figure}
We describe the decreasing mass of a bound cluster that survived infant
mortality ($t \simgreat 10^7$ yrs) as a function of time.
 Let us define $\Mpt$ as the mass of a cluster of initial mass
$\Mi$ and age $t$, and $\mut=\Mpt/\Mi$ as the fraction of the initial 
mass that is
still in the cluster. The  decrease of mass due to both stellar evolution
and disruption can then be described as

\begin{equation}
\frac{d\Mp}{dt}= \left(\frac{d\Mp}{dt}\right)_{\rm ev} +
                  \left(\frac{d\Mp}{dt}\right)_{\rm dis}
\label{eq:dmpdt}
\end{equation}
where the first term describes the evolution by stellar mass loss and
the second term by disruption.
Following the arguments given in Sect. 2 we assume that we can
describe the mass loss by disruption as

\begin{equation}
\left(\frac{d\Mp}{dt}\right)_{\rm dis} = \frac{-\Mp}{\tdis} = 
               \frac{-\Mp}{\tgal (\Mp/\Msun)^{\gamma}}= 
               \frac{-\Msun}{\tgal}\left( \frac{\Mp}{\Msun}\right)^{1-\gamma}
\label{eq:dmpdtdis}
\end{equation}
with $\gamma=0.62$ and $\tgal$ is a constant  
that depends on the tidal field. (See Lamers, Gieles \& Portegies
Zwart, 2005 for the
dependence of $\tgal$ on the conditions in different galaxies).
The first equality assumes that the mass lost by disruption can be
approximated by an exponential decay with a time scale that
decreases as the mass of the cluster decreases. 
This is equivalent to
the statement that the disruption time in our description is {\it defined} as
$\tdis^{-1}=d \ln \Mp /dt$. 
The second equality assumes that this timescale depends on the
mass as $\Mp^{\gamma}$. (If $\gamma$ was equal to 1 
and the evolutionary mass loss could be ignored, then
the mass of the cluster would decrease linearly with time until 
$t=\tgal \Mi/\Msun$. For $\gamma=0$ the decrease would be exponential.) 

Equation \ref{eq:dmpdt} 
can easily be solved numerically. 
It turns out that the mass decrease of a cluster can be 
approximated very accurately by the following formula

\begin{equation}
\mu(t;\Mi)\equiv \frac{M(t)}{M_i}\simeq 
\left\{(\muevt)^{\gamma}-
\frac{\gamma t}{\tgal}\left(\frac{\Msun}{\Mi}\right)^{\gamma}
\right\}^{1/\gamma}
\label{eq:muapprox}
\end{equation}
if the first term in brackets is larger than the second term.
If the second term is larger than the first term, i.e. when the mass
lost by disruption is larger than the mass that remained
after mass loss by stellar evolution, then  $\mut=0$
and the cluster is completely disrupted. 
Approximation \ref{eq:muapprox} is quite accurate because during the 
first $10^8$ years mass loss is dominated by stellar evolution 
so the second term is negligible and \muevt\ describes the fraction
of the mass that survives mass loss by stellar evolution.
During later years, when $\muevt$ decreases very slowly, 
the mass loss is dominated by disruption, which is described by the
second term in brackets. 
Eq. \ref{eq:muapprox} can be inverted to express the initial cluster
mass in terms of the present cluster mass:

\begin{equation}
M_i \simeq 
\left\{\left(\frac{\Mp}{\Msun}\right)^{\gamma}+
\frac{\gamma t}{\tgal}\right\}^{1/\gamma} \muevt^{-1}
\label{eq:miapprox}
\end{equation}

Figure \ref{fig:1} compares the numerical solution with the analytic 
approximation for  various initial masses, $10^3 \le \Mi /\Msun \le 10^6$, 
for a short and a long disruption timescale, $\tgal=2$ Myr and 30 Myr, 
both for Z=0.02.
In all cases the agreement between the analytic and the numerical 
solution is excellent, i.e. within about 0.015 dex, although the
disruption times vary by more than 6 orders of magnitudes from model
to model.
Even for the low mass cluster model of $\Mi=10^3$ \Msun\ at
$\tgal=2$ Myr, for which the disruption is already effective during
the first 10 Myr, the agreement between the the numerical solution
and the analytic expression is very good.
(Tests show that Eq. \ref{eq:muapprox} is also a very good
approximation for all other values of $\gamma$ in the range of $0<\gamma<1$.)

We define $\ttot$ as the total disruption time and \tdelta\ as the
time when only a fraction $\Delta$ of the initial mass remains.
>From Eq. \ref{eq:muapprox} 
we find that \ttot\ and \tdelta\ are described by the implicit relations

\begin{equation}
\ttot=\frac{\tgal}{\gamma} \left(\frac{\Mi}{\Msun}\right)^{\gamma} 
\{\muevol(\ttot)\}^{\gamma}
\label{eq:ttot}
\end{equation}
and

\begin{equation}
\tdelta=\frac{\tgal}{\gamma} \left(\frac{\Mi}{\Msun}\right)^{\gamma}
(\{\muevol(\tdelta)\}^{\gamma}-\Delta^{\gamma})
\label{eq:t95}
\end{equation}
We will use this last expression for a comparison of our analytic
solution with those of \nbody-simulations. 
In the range of $10^4<\tgal<10^7$ years and $10^3 <\Mi <
10^6$ \Msun\ the values of \ttot\ can be approximated  by 

\begin{eqnarray}
\log(\ttot) & \simeq & \log \left(\frac{\tgal}{\gamma}\right)
+ \gamma \log \left(\frac{\Mi}{\Msun}\right) \nonumber \\
 & &  - 0.00825~ \log \left(\frac{\Mi}{\Msun}\right)
\times \log\left(\frac{\tgal}{10^4 {\rm yr}}\right)
\label{eq:ttotapprox}
\end{eqnarray}
This approximation is valid within 0.03 dex for all metallicities.
For all models we find that $\tninefive \simeq 0.89~\ttot$. 

These equations imply that the total disruption time of a cluster
with an initial mass of $10^4$ \Msun, which was referred to as $t_4$ in
BL03 and in Lamers, Gieles \& Portegies Zwart (2005), is related to \tgal\ by

\begin{equation}
t_4^{\rm total} = \frac{1.355~ 10^{4 \gamma}}{\gamma} \tgal^{0.967} 
    = 6.60\times 10^2 ~\tgal^{0.967}
\label{eq:t0t4}
\end{equation}
where the last equation is only valid if $\gamma=0.62$ and \tgal\ is
in yrs. 

%
Equation \ref{eq:t0t4} shows  that $\ttot(\Mi)$ is approximately 
 proportional to
$\Mi^{\gamma}$, which was the relation adopted in the study of the
disruption times of clusters in different galaxies by
BL03 and Lamers, Gieles \& Portegies Zwart (2005).

\section{Comparison of the analytic solution with results 
of $N$-body simulations}

\begin{figure*}
\centerline{\psfig{figure=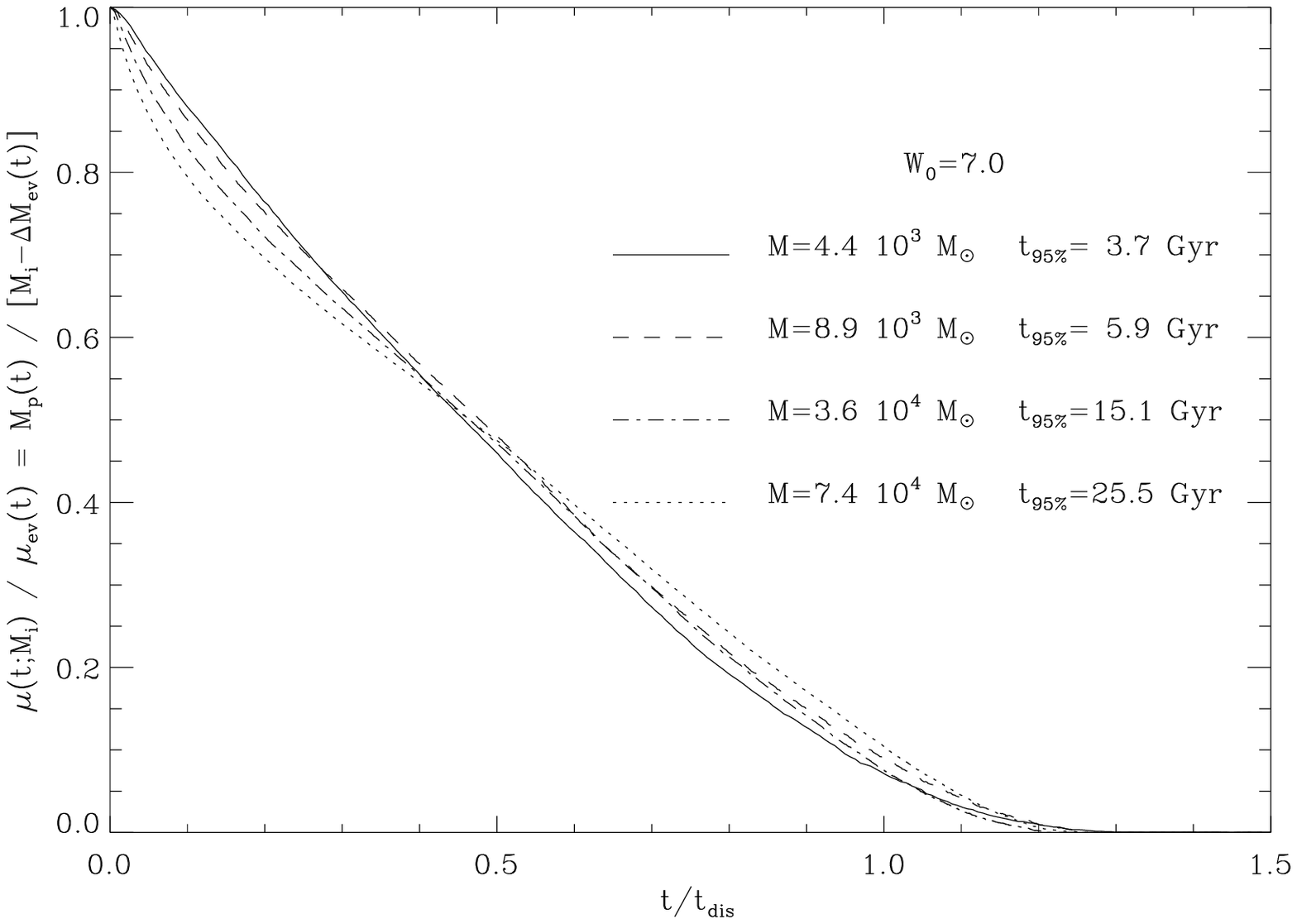,width=9.4cm}
            \psfig{figure=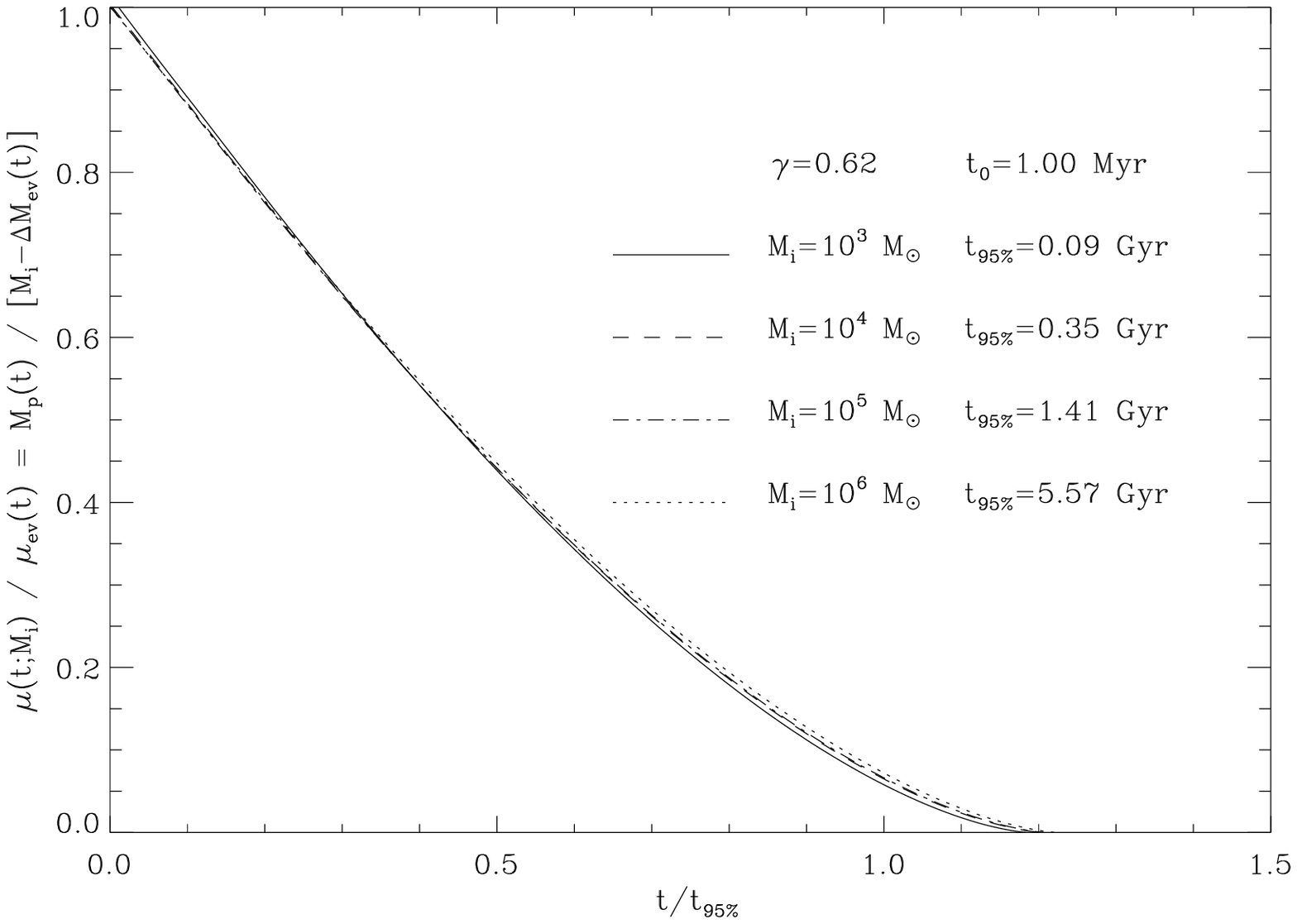,width=9.4cm}}
\centerline{\psfig{figure=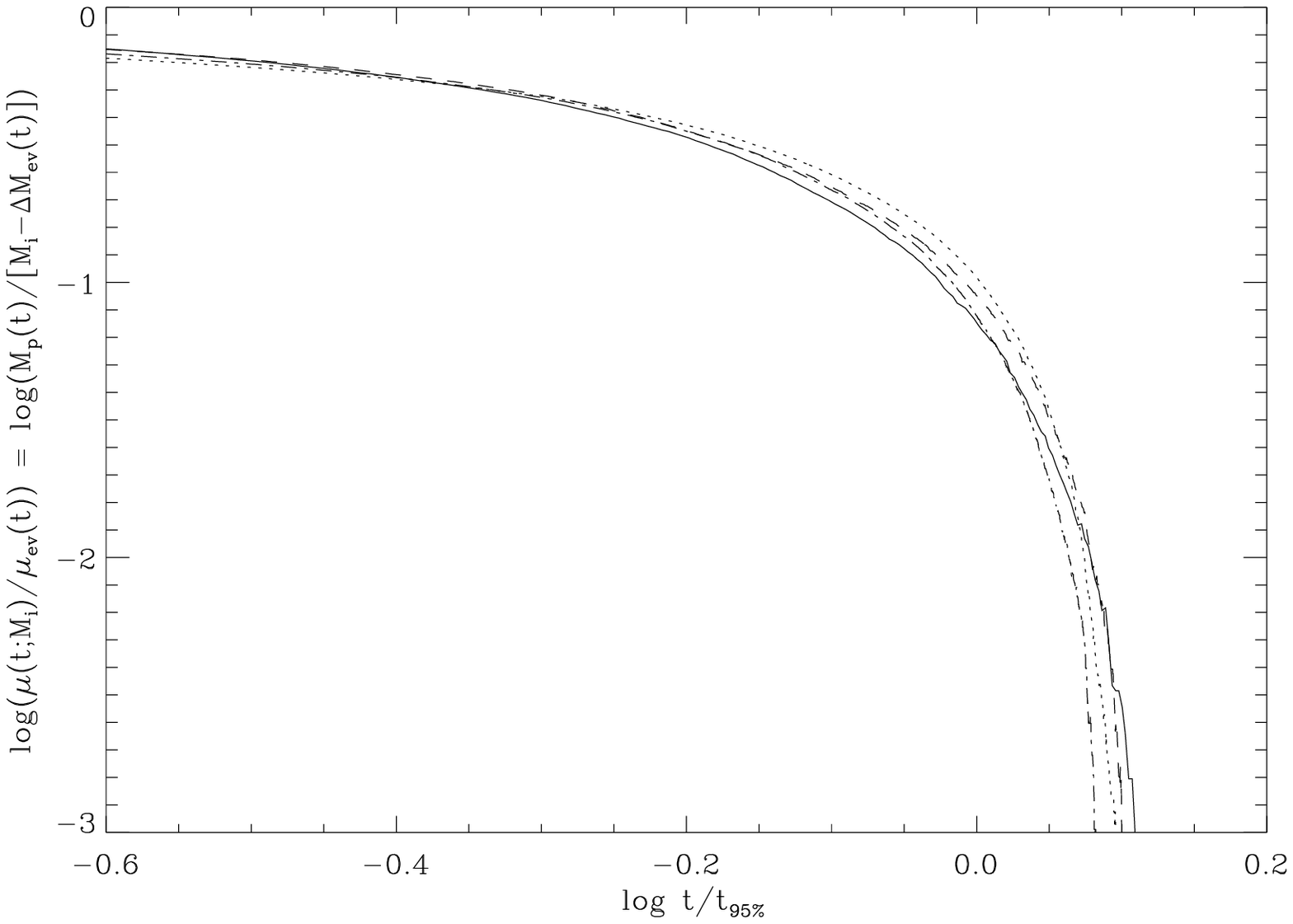,width=9.4cm}
            \psfig{figure=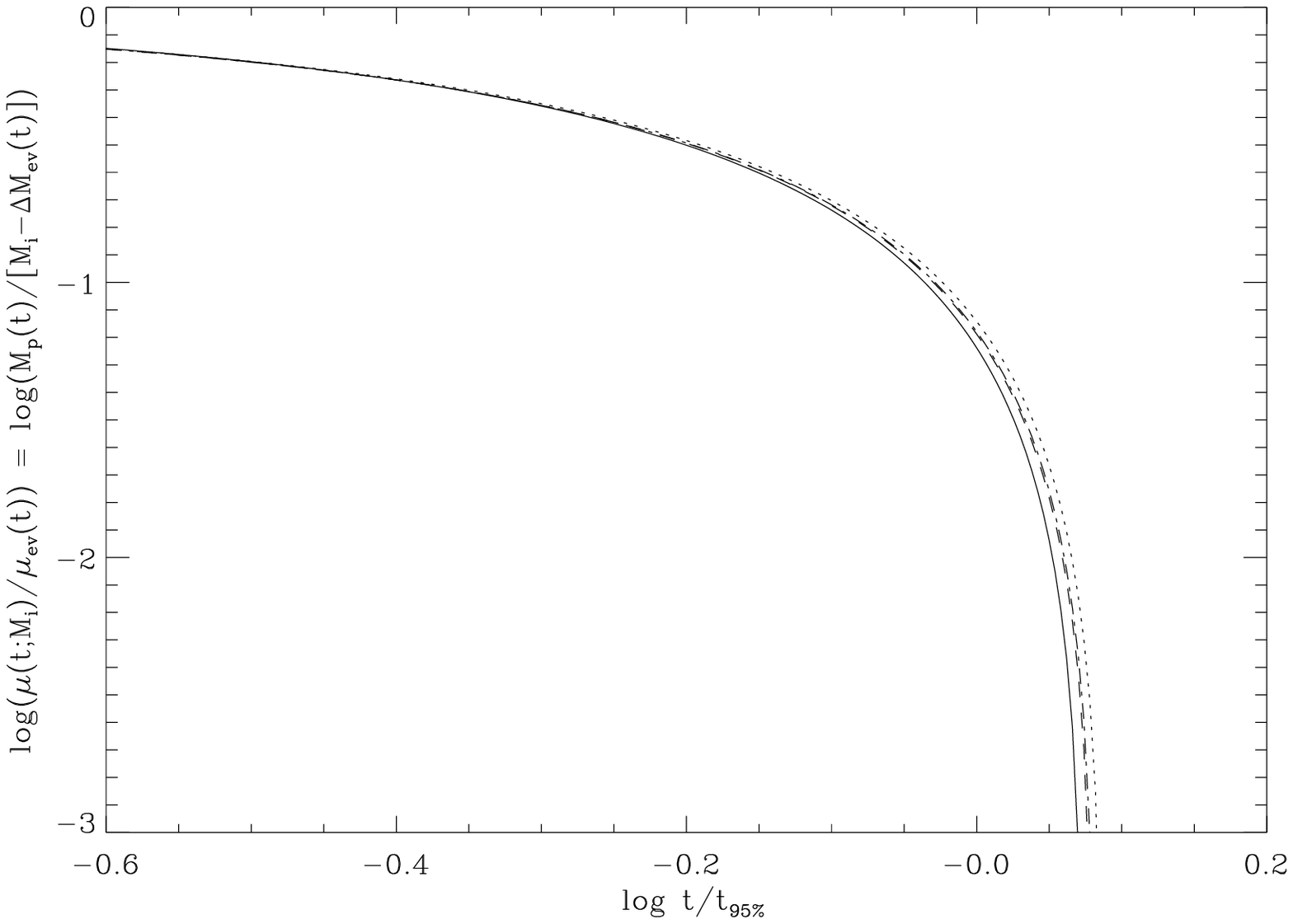,width=9.4cm}}
\caption[]{Comparison of the decrease of the cluster mass with
            time between the results of the $N$-body simulations 
by BM03 (left) and our description (right) for clusters of different
initial numbers of stars, $N_{\rm i}$ or mass $M_{\rm i}$. The mass
has been corrected to first order for the mass lost by stellar evolution.
In all figures the time $t$ is scaled to $\tninefive$, i.e. 
the time when 95\% of the cluster mass is lost due to stellar
evolution and disruption. 
The upper figures are for mass versus time and the lower figures are
for log(Mass) versus log(time). Notice the strong similarity between
the results of the $N$-body simulations and our simple description
(Eq. 6).}
\label{fig:comparison}
\end{figure*}


We can compare our analytic expression for the decreasing mass
of a cluster with the results of $N$-body simulations.
We adopt the simulations by BM03 who calculated
the fate of clusters under various conditions. 
Before making this comparison, we would like to point out that:\\
(a) BM03 adopted the stellar IMF of Kroupa (2001), which have an
initial mean mass of 0.547 \Msun, whereas 
our calculations are based on the GALEV cluster evolution models
(see Sect. 3.1) which have a Salpeter (1955) IMF with a lower mass 
cut-off of 0.15 \Msun\ and a maximum mass of 85 \Msun,  
resulting in an initial mean mass of 0.516 \Msun. \\
(b) BM03 adopted the evolutionary mass loss rates from Hurley, Pols
\& Tout (2002), whereas those of the GALEV models are based on the
calculations from the Padova-group (see Sect. 3.1).\\
(c) BM03 define the disruption time of a clusters, $\tdisBM$, (called
dissolution time in their paper) as the time at which only 5 percent of the 
initial mass is still in the cluster. \\
(d) The simulations by BM03
show that the mean mass of the remaining stars in a dissolving cluster 
changes with time. In the early phase the mean mass decreases 
because stellar evolution removes the massive stars, 
but the mean mass increases in later phases when disruption becomes
the dominant mass loss mechanism and low mass stars are lost preferentially.
In the simulations by BM03 the mean mass near the end of the life 
of a cluster of initially $3\times 10^5$ stars has increased from 0.516 to 1.2
\Msun.
In our analytic approximation this effect is not taken into account
Therefore we can expect a small offset in the timescale between
the \Nbody\ and the analytic results due to mass segregation.\\

Because of differences in the adopted mass loss by stellar evolution
between our and BM03 models, 
we compare the results for the decreasing mass of a cluster not
directly, but corrected for the stellar evolution. This means that we
compare the predictions for $\mu(t;\Mi)/\muevol(t)$ rather than for
$\mu(t;\Mi)$. The function $\mu(t;\Mi)/\muevol(t)$
describes the fraction of the initial mass of the cluster that is lost by
disruption only. This function, which is called $M_{\rm rel}(t)$ by
BM03, is expected to be approximately 
independent of the adopted evolutionary mass loss rates.

BM03 give the function $M_{\rm rel}(t)$ 
for their models of clusters of different
initial numbers of stars, $8.2\times 10^3<N_{\rm i}<1.3\times 10^5$, 
and with an
initial concentration described by  $W_0=7$, in circular
orbits around the galactic center at a distance of 8.5 kpc. 
(see BM03 Fig. 6). Their results show that $M_{\rm rel}(t)$
decreases almost linearly with time, and that the function plotted
against $t/\tdisBM$ is about the same for all clusters.
We compare their results with our analytic expression Eq. \ref{eq:muapprox}.

\begin{figure}
\centerline{\psfig{figure=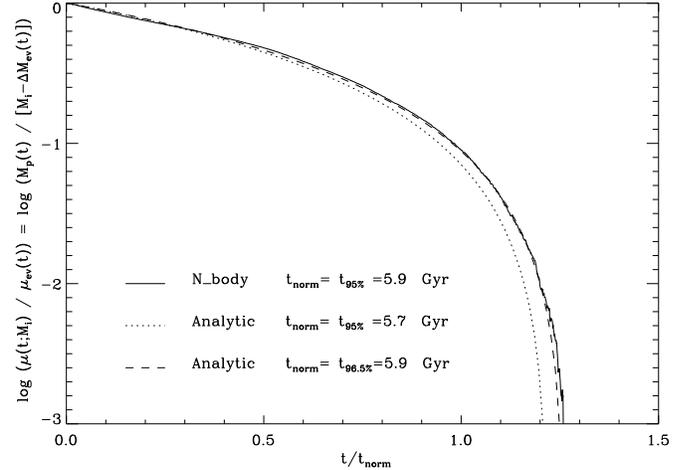,width=9.4cm}}
\caption[]{Comparison between the mass decrease of a cluster of 
$8.9\times 10^3M_{\odot}$ predicted by $N$-body simulation of BM03 (full
line) and our analytic approximation for 
$\tgal=18$ Myr (dashed and dotted line). 
The mass decrease has been corrected for the
mass lost by stellar evolution. The timescale is normalized to the
time when 95 or 96.5\% of the initial mass is lost (see text).}
\label{fig:comparison2}
\end{figure}

The left hand panels of Fig. \ref{fig:comparison} 
show the normalized decrease in cluster 
mass, $M_{\rm rel}(t/\tdisBM)$, of the models by BM03.
The top figure gives the mass as function of time, both in linear
scale. However, since the
mass of clusters will decrease orders of magnitudes before they are 
disrupted, we also plot the logarithm of $M_{\rm rel}(t)$
as a function of the logarithm of $t/\tdisBM$ in the lower figure. 
The right hand panels of Fig. \ref{fig:comparison} show
the results of our models. For a fair comparison between our 
calculations and those of BM03, we plot $\mu(t;\Mi)/\muevol(t)$ as a
function of $t/\tninefive$.  The
agreement between the predictions of the $N$-body calculations and
our description is very good, both in linear scale as in
logarithmic scale. The very small difference of less than 5 percent
during the early phase of the most massive models is probably due to the 
difference in the adopted mass loss by stellar evolution.
(The plotted relations are to first order corrected for the effects of
stellar evolution. However because the cluster evolution is caused by
the combination of stellar evolution and tidal effects, the corrected 
relations still bear the imprint of the adopted stellar evolution.)
Figure \ref{fig:comparison2} shows the direct comparison between the
\Nbody\ prediction for a clusters of initial mass $M_i=8.9\times 10^3 \Msun$
($N_i=16384$) with a concentration of  $W_0=7$ and our analytic
solution. For the analytic solution we adopted the same mass and 
the parameter $\tgal=18$ Myr was chosen in such a way that the
95 \% disruption time is 5.7 Gyr, very close to that of the \Nbody\
simulation.
The times are normalized to the time when $95\%$ of the clustermass is
gone. We see that the prediction by the \Nbody\
calculations (full line) and the analytic expression (dotted line) are
very similar, apart from a small offset of the timescale.
If we normalize the timescale of the analytic solution to the time
when $96.5$\% of the initial mass is gone (dashed line), the agreement
is almost perfect! This difference in timescale is due to the fact that in
the  \Nbody\ simulations there is a preference for the low mass stars
to be kicked out of the clusters, whereas in the analytical solution
stars of all masses are lost. Apart from this small difference, the
analytical solution with \tgal\ as a free parameter describes the 
decrease of the mass of clusters surprisingly accurately.
 
\subsection{Explanation}

We conclude that our analytic description of the decreasing
mass of a cluster, with the adjustable free parameter \tgal, 
is very similar to the one derived by $N$-body
simulations. We want to stress that this result 
is not trivial. 

The values of $\tdisBM$
that results from the \nbody\ simulations by BM03
depend on the initial number of stars as $N_{\rm i}^{0.62}$, in
excellent agreement with the empirically derived dependence 
by BL03 (see Sect.2).
However this does not automatically imply that the decrease
of mass with time of {\it each individual cluster} can be described by
a function that depends on the $\it present$ mass of that cluster 
to the same power $\gamma=0.62$ (Eq. \ref{eq:dmpdtdis}).  
However the good agreement between the simulations by 
BM03 and our result from the analytic description shows that it does.

 So, not only does the total lifetime of all clusters depend on their 
initial mass as $\ttot \propto \Mi^{0.62}$, but also at every
moment during the lifetime of a cluster the exponential disruption time,
$(d \ln \Mp/dt)^{-1}$, scales with the mass to the same power
0.62. This is a consequence of the fact that the timescale of the
disruption process depends on both the half mass relaxation time,
$t_{\rm rh}$ and the crossing time $t_{\rm cr}$ as about
$t_{\rm rh}^x t_{\rm cr}^{1-x}$, with $x \simeq 0.75$ for 
models with a concentration parameter $W_0=5.0$ and $x=0.82$
if $W_0=7.0$ (BM03). For both sets of models the disruption time
scales with mass as $M^{0.62}$ (Gieles et al. 2004). 
It is the continuous adjustment of the half mass
relaxation time and the crossing time to the changing conditions 
of the cluster that results in an exponential disruption time that
varies during the life of a cluster as the {\it present} mass
$\Mp^{0.62}$.


\section{The predicted mass and age distributions of cluster samples}
\label{sec:5}

Using the expression for the decreasing mass of clusters,
Eq. \ref{eq:muapprox}, we can predict the mass and age distribution 
of cluster samples (for open clusters as well as globular clusters) 
for any adopted cluster formation rate, $CFR(t)$, 
and cluster initial mass function, CIMF. 

 Suppose that the CIMF is a power law with a slope $-\alpha=-2$
(Zhang \& Fall 1999; Larsen 2002; Bik et al. 2003; de Grijs et
al. 2003) in the range of $\Mmin < M_{\rm cl} < \Mmax$, 
$\Mmin \approx 10^2 \Msun$ and $\Mmax \approx 10^7 \Msun$,
then the 
number of clusters with initial mass \Mi\ formed at time $t$ will be

\begin{equation}
\nmit~=~S(t) \left( \frac{M_i}{\Msun}\right)^{-\alpha} 
~{\rm for}~~\Mmin < M_i < \Mmax
\label{eq:cimf}
\end{equation}
in Nr $\Msun^{-1}$yr$^{-1}$ if
$t$ is in years and $M_i$ in \Msun. 
The function $S(t)$ is related to the cluster formation rate $CFR(t)$
in Nr~yr$^{-1}$ as

\begin{eqnarray}
\cfr &=& \int^{\Mmax}_{\Mmin} \nmit dM_i \nonumber\\
 &=&
\frac{S(t)}{1-\alpha} 
\left\{ \left( \frac{\Mmax}{\Msun}\right)^{1-\alpha}
  - \left(\frac{\Mmin}{\Msun}\right)^{1-\alpha} \right\} \nonumber\\
\label{eq:cfra}
\end{eqnarray}
The total mass of the clusters formed per year  
is $S(t)  \ln(\Mmax / \Mmin)$ in $\Msun$~yr$^{-1}$ for $\alpha=2$.


\subsection{The distribution of the masses and ages}
\label{sec:5.1}

Given the initial mass distribution of the clusters, their formation
rate, $CFR(t)$, and an expression for the way in which the mass of each
cluster changes with time (Eq. \ref{eq:muapprox}), we can calculate
the present distribution of existing clusters as a function of age
or as a function of their mass.

If $N(\Mp,t)$ is the number of clusters
of mass \Mp\ and age $t$ in Nr~\Msun$^{-1}$~yr$^{-1}$,
then $N(\Mp,t)$ and $N(\Mi,t)$ are related by the conservation of the
numbers of clusters 

\begin{equation}
N(\Mp,t)~d\Mp  = N(\Mi,t)~d\Mi  
\label{eq:dMp}
\end{equation}
with $\Mp(t)$ and $\Mi(t)$ related via Eq. \ref{eq:muapprox}. Applying
the derivative $d(\Mp,t)/d(\Mi,t)$, that follows from
Eq. \ref{eq:muapprox}, and combining this with Eq. \ref{eq:cimf}
for the CIMF and Eq. \ref{eq:cfra} for the CFR 
we find the present distribution of clusters
as function of mass and age:

\begin{eqnarray}
N(\Mp,t) &=& S(t) \left(\frac{\Mp}{\Msun}\right)^{-\alpha}
\muevt^{\alpha-1} \nonumber \\
& & \left\{ 1+\frac{\gamma t}{\tgal}
  \left(\frac{\Mp}{\Msun}\right)^{-\gamma}
\right\}^{(1-\alpha-\gamma)/\gamma}
\label{eq:dnmpt}
\end{eqnarray} 
%
This equation is valid for $\Mp$ smaller than  some upper limit, 
\Mptmax, which is 
the mass
of a cluster of age $t$ with the maximum initial mass $\Mmax$   

\begin{equation}
\Mptmax= \Mmax \left\{(\muevt)^{\gamma}-
\frac{\gamma t}{\tgal}\left(\frac{\Msun}{\Mmax}\right)^{\gamma}
\right\}^{1/\gamma}
\label{eq:Mptmax}
\end{equation}
%
Similarly, for a given value of $\Mp$, Eq. \ref{eq:dnmpt} is only valid 
for ages less than $\tupmp$ which is the age at which a cluster with
an initial mass of \Mmax\ has reached a mass \Mp. So $\tupmp$ is
given by the condition

\begin{equation}
\left(\frac{\Mp}{\Mmax}\right)^{\gamma}
+ \frac{\gamma \tup}{\tgal}\left( \frac{\Msun}{\Mmax} \right)^{\gamma}
-(\muevol(\tup))^{\gamma} = 0
\label{eq:tmaxmp}
\end{equation}
Eq. \ref{eq:dnmpt} allows us to calculate the predicted mass and age 
distribution of a cluster sample for any assumed cluster formation
rate. 
The mass distribution is found by integrating 
$N(\Mp,t)$ over age for any mass, and the age distribution is found
by integrating over mass between $M_{\rm up}(t)$ and some lower mass
limit $M_{\rm low}(t)$, set by the detection limit, for any age.

\begin{figure}
\centerline{\psfig{figure=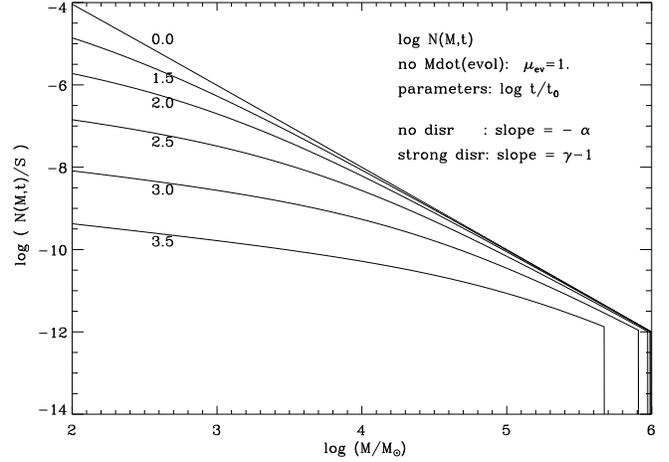,width=9.2cm}}
\caption[]{The changes in the mass distribution (Eq. 15) of 
a sample of clusters as a function of their age, in case
stellar evolution can be neglected. We adopted a cluster
initial mass function in the range of $10^2<M<10^6~\Msun$ with
$\alpha=2.0$ and a disruption parameter $\gamma=0.62$. 
The different curves refer to different ages, which are
parametrized by log $t/\tgal$. The maximum
mass decreases with age due to disruption.}
\label{fig:Nmt}
\end{figure}

\begin{figure}
\centerline{\psfig{figure=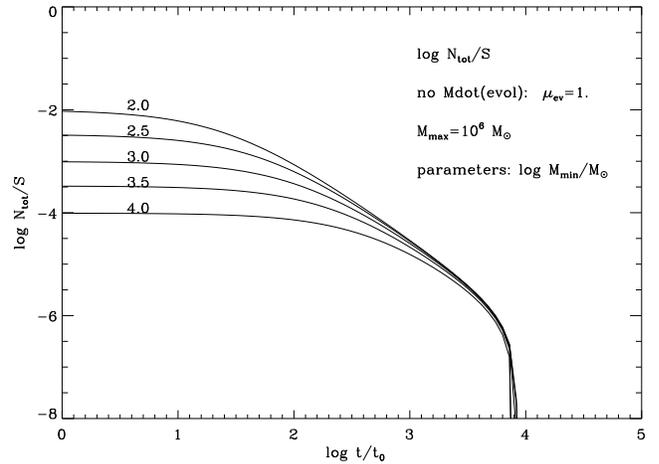,width=9.2cm}}
\caption[]{The age distribution of cluster samples formed at a
  constant formation rate, in case
stellar evolution can be neglected. We adopted a cluster
initial mass function in the range of $M_{\rm min}<M<M_{\rm max}$ 
with $M_{\rm max}=10^6 \Msun$ and different values of $M_{\rm min}$.
We adopted $\alpha=2.0$ and a disruption parameter $\gamma=0.62$. 
The curves are labeled with  log $(M_{\rm min}/\Msun)$.}
\label{fig:Ntot}
\end{figure}

The mass and age distribution of cluster samples 
$N(M,t)$ depends on the stellar evolution and on disruption. 
To get insight into the effect of disruption on the evolution of a
cluster sample we first consider a simplified case when mass loss by
stellar evolution is neglected, i.e. $\mu_{\rm ev}(t)=1.0$.
In that case the function
$N(M,t)/S(t)$ depends only on the slope $\alpha$ of the cluster IMF, 
the mass-dependence  $\gamma$ of the disruption and on
the ratio $t/\tgal$. Figure \ref{fig:Nmt} shows the shape
of $N((M,t)/S(t)$.  For very young ages
or very long disruption time ($t/\tgal \le 10 $) the distribution
is the initial CIMF with slope $-\alpha$. 
For strong disruption, i.e. $t/\tgal \ge 10^2$,
the distribution of  the low mass clusters becomes flatter and 
approaches a power law of the type $N(M)\sim M^{\gamma-1}$.
This distribution is similar to the one predicted by BL03 
for instantaneous disruption, except that the transition
between the two slopes is gradual, whereas it shows a sharp kink
 for models with instantaneous disruption.

Figure \ref{fig:Ntot} shows the normalized age distribution 
$N_{\rm tot}(t)/S$, for cluster samples
formed at constant cluster formation rate in the mass range of $M_{\rm
  min} < M <M_{\rm max}$ for various values of $M_{\rm min}$, 
in case the mass loss by stellar evolution
can be ignored. The distribution is flat for young clusters
at a value of $N/S \simeq M_{\rm min}^{-1}$
and curves down to older clusters, approaching a slope 
$N \sim (t/\tgal)^{-1/\gamma}$. This was predicted by BL03 for instantaneous disruption. The distribution drops to
zero at the age at which the most massive clusters are disrupted,
i.e. when $t/\tgal = M_{\rm max}^{\gamma}/\gamma$ (Eq. 16), which is at
$t/\tgal=8.46~10^3$ for $M_{\rm max}=10^6 \Msun$ and $\gamma=0.62$.


\section{Application to Galactic open clusters}
\label{sec:open}

Ideally one would like to compare the predictions with complete (or at least
unbiased) samples of clusters with known masses and ages. Unfortunately
this is not possible at the moment, because samples of clusters in
external galaxies are usually magnitude limited. (The method for
determining the disruption times from magnitude limited cluster
samples with gradual disruption will be described by Lamers (2005) and
applied to the cluster sample in M51 by Gieles et al. (2005a)).
Samples of open clusters in the solar neighbourhood are unbiased,
but only the cluster ages have been determined systematically 
and not the cluster
masses. We will compare our predictions for the age distribution
to the sample of open clusters in the solar neighbourhood.


\begin{figure}
\centerline{\psfig{figure=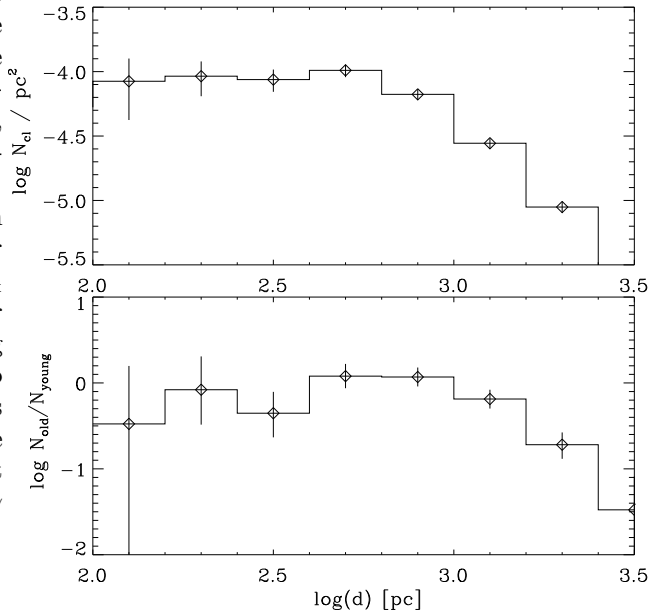,width=12.0cm}}  
\caption[]{Top: the surface density distribution projected onto the
  Galactic plane of open clusters
in the solar neighbourhood from the homogeneous Kharchenko et al. (2005) 
catalogue. Errorbars indicate 1$\sigma$ statistical uncertainties.
The surface density is about constant up to at least 600 pc and
possibly 1 kpc. Bottom: the ratio between the numbers of old
($>2.5~10^8$ yr) and young ($<2.5~10^8$ yr) clusters as a function of
distance. The ratio is about constant up to 1 kpc.
}
\label{fig:distance}
\end{figure}

\begin{figure}
\centerline{\psfig{figure=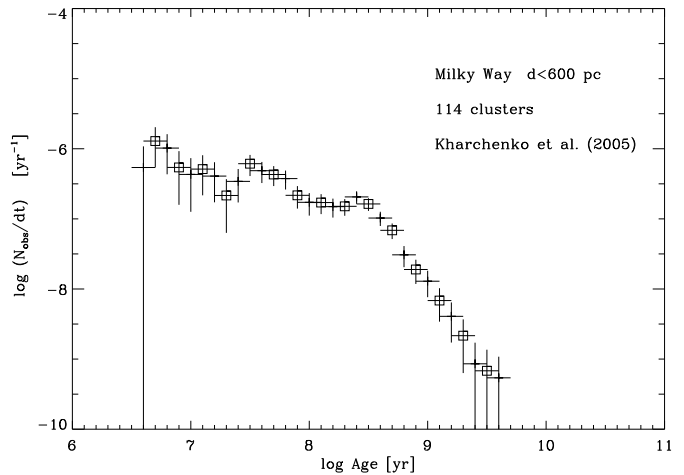,width=9.4cm}}
\caption[]{ The age histogram in units of number per
year, in logarithmic age-bins of 0.2 dex,
of 119 open clusters 
within $d<600$ pc  from Kharchenko et al. (2005). 
The
distributions are plotted for two sets of bins, shifted by 0.1 dex,
with and without squares respectively. The error-bars indicate the 
1$\sigma$ statistical uncertainty. The distribution decreases to older ages,
with a small bump around $\log (t/{\rm yr}) \simeq 8.6$. For 
$\log (t/{\rm yr}) < 7.5$ the distribution is uncertain due to large errorbars.
}
\label{fig:kharchenko_age}
\end{figure}


\subsection{The sample of open clusters}

Kharchenko et al. (2005) have recently published a catalogue of
astrophysical data of 520 galactic open clusters (COCD = Catalogue of
Open Cluster Data) in the wider
neighbourhood of the Sun with the values of angular sizes of cluster
cores and coronae, heliocentric distances $d$, $E(B-V)$, mean proper
motions, radial velocities and ages. These parameters have been
determined by homogeneous methods and algorithms including a careful procedure
of cluster member selection. The basis of this study is the ASCC-2.5
- All-Sky Compiled Catalogue of about 2.5 million stars (Kharchenko
2001) down to $V \simeq 14$ (completeness limit at $V \simeq 11.5$),
with compiled proper motions and $B,V$ magnitudes, which is based on
the $Tycho-2$ data, and supplemented with $Hipparcos$ data sets, as well
as with some ground-based catalogues
\footnote{The ASCC-2.5 catalogue can be retrieved from the CDS at 
  ftp://cdsarc.u-strasbg.fr/pub/cats/I/280A}.
Cluster membership is based on a combined probability which takes into
account kinematic (proper motion), photometric, and spatial
selection criteria (see Kharchenko et al. 2004 for details). For
stars within a circle with a cluster radius the membership probability
is calculated as a measure of a deviation
either from the cluster mean proper motion (kinematical probability),
or from the Main Sequence edges (photometric probability). Stars
deviating from the reference values by less than one $\sigma$ (rms)
are classified as most probable cluster members (1$\sigma$-members,
i.e., with a membership probability $P \ge 61$\%). Those falling in
semi intervals [1$\sigma$,2$\sigma$) or [2$\sigma$,3$\sigma$) are
considered as possible members ($P=14-61$\%) or possible field stars
($P=1-14$\%), respectively. Stars with deviations larger than
3$\sigma$ are regarded as definite field stars ($P < 1$\%). As a
rule, all cluster parameters were determined from the data of the
most probable cluster members. Cluster ages were determined with an
isochrone-based procedure which provides a uniform age scale
(see Kharchenko et al. 2005 for details). Thus the COCD is the most
homogeneous and most complete catalogue of open clusters in the solar
neighbourhood available up to now.

Figure \ref{fig:distance} shows the distance  distribution
of the density of clusters projected onto the
Galactic plane, in number per pc$^2$. We see that
within the statistical uncertainty the surface density is constant 
up to at least 600 pc, and possibly even up to 1 kpc.
The lower part of the figure shows the ratio between old ($ t>
2.5~10^8$ yr) and young ($ t< 2.5~10^8$ yr) clusters as a function of
distance. Up to a distance of about 1 kpc there is no
significant change in this ratio within the statistical uncertainty.
This is important for our study, because it shows that the age
distribution of open clusters within about 1 kpc is not affected by
detection limits.

Figure \ref{fig:kharchenko_age} shows the age distribution 
in number per year of the 114 clusters within 600 pc in the Kharchenko
et al. (2005) sample. 
The effect of binning is demonstrated by plotting two sets of data,
where the bins have been shifted by 0.1 dex relative to one another.
This distribution is decreasing with age, apart from a small
local maximum around $\log(t/{\rm yr}) \simeq 8.5$. The distribution
at young ages is sensitive to the choice of the age-bins and shows a
significant scatter. The steep slope at $\log(t/{\rm yr}) > 8.8$
demonstrates that cluster disruption is important.


\subsection{The lower mass limit of the clusters}

For the determination of the cluster formation rate and the disruption
times we need an estimate of the minimum mass of the clusters in the 
Kharchenko et al. (2005) sample. This catalogue does not list  
the mass of the clusters, but it can be estimated roughly
from the age, distance, extinction and the number of stars of each cluster.
We have estimated the lower mass limit of the Kharchenko et al. cluster
sample in the following way. 

(a) First we calculate the number of members brighter than the
completeness limit, $V_{\rm lim} = 11.5$, within the cluster radius with the
following constraints on cluster membership probabilities:
$2\sigma$ photometric probability and $2\sigma$ kinematic probability.
These probabilities were defined in
Sect 6.1. We did this separately for main sequence stars only, 
and for members of all spectral types.\\
(b) Using the distance and
$E(B-V)$ of the clusters, we expressed $V_{\rm lim}$ in $M_v$.
This limiting magnitude of the main sequence (MS) stars is expressed in
$M_{\rm bol}$ and mass, $M^*_{\rm lim}$, using the 
bolometric corrections and the
mass luminosity relation of luminosity class V stars. 
With the cluster age known, the
mass of the stars at the turn-off point of the MS, 
$M_{\rm TO}$, can be estimated from the relation between the MS
lifetime and stellar mass, for which we adopted the relation by
Schaller et al. (1992) for solar metallicity.\\
(c) We then assumed a stellar IMF with a slope of -2.35,
i.e. $N(M)dM=C M^{-2.35}dM$, and calculated the value of $C$ that gives the
derived number of main sequence stars in the mass range 
of $M^*_{\rm lim} <M < M_{\rm TO}$.\\
(d) With this value of $C$ we calculated the total mass of the cluster
for all stars between  the upper MS mass limit, $M_{\rm TO}$,
and a lower limit for the stellar mass, $M^*_{\rm min}$, 
for which we adopted $0.15$\Msun. (With this lower mass limit the mean
stellar mass of a cluster with a Salpeter mass function is $0.51
\Msun$, which is quite similar to the mean mass of $0.55 \Msun$ for a Kroupa
(2001) mass function.)    
We corrected this mass for the small number of stars 
with a MS age that is 15\% shorter
than the age of the cluster, in order to correct for the stars
that have evolved off the MS but have not yet ended their live. 
So the adopted mass range is 
$M^{\rm max}_{\rm alive} < M < M^*_{\rm min}$, where 
$M^{\rm max}_{\rm alive}$ is the mass of a star with a MS lifetime of
0.85 times the age of the cluster.
(We note that white dwarfs, neutron stars and black holes 
do not add significantly
to the mass of clusters with ages less than about a few Gyr.)
In this way we estimated the mass of all the clusters in the
Kharchenko sample with $d<600$ pc.\\
(e) We also applied this method directly to the observed 
number of the probable (2$\sigma$) member stars with
$V<11.50$, of all luminosity class. These are the
observed stars in the mass range of 
$M^{\rm max}_{\rm alive} < M < M^*_{\rm min}$.
The resulting masses are
very similar to those derived from the number of MS stars only, 
except for a few clusters of high extinction for which the
$V_{\rm lim}=11.5$ corresponds to stars near the top of the main sequence
and the number of probable members brighter than $V=11.5$ is small.\\
(f) To estimate the sensitivity of the resulting cluster mass to the
adopted stellar lower mass limit, we repeated the analysis for an
adopted lower mass of $M^*_{\rm min}=0.25 \Msun$. In this case the
estimated masses are about 80\% of those estimated for 
$M^*_{\rm min}=0.15 \Msun$.

\begin{figure}
\centerline{\psfig{figure=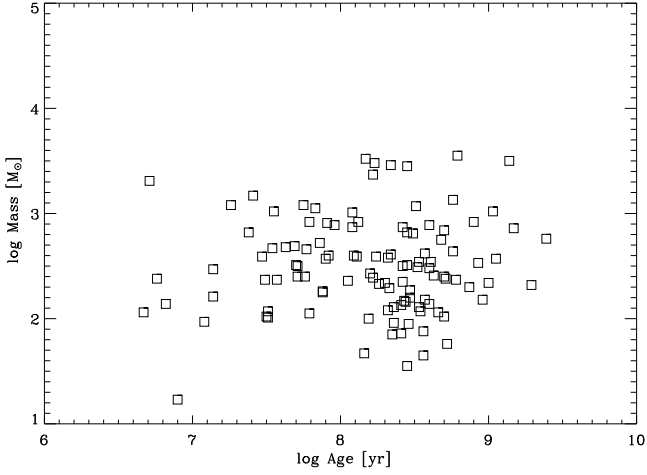,width=9.4cm}}  
\caption[]{
The mass-versus-age diagram of 114 clusters of the Kharchenko et
al. (2005) catalogue within a distance of 600 pc. The mass is derived
from the number of main sequence stars with $V<11.50$.}
\label{fig:mass_age_hist}
\end{figure}
 
\begin{figure}
\centerline{\psfig{figure=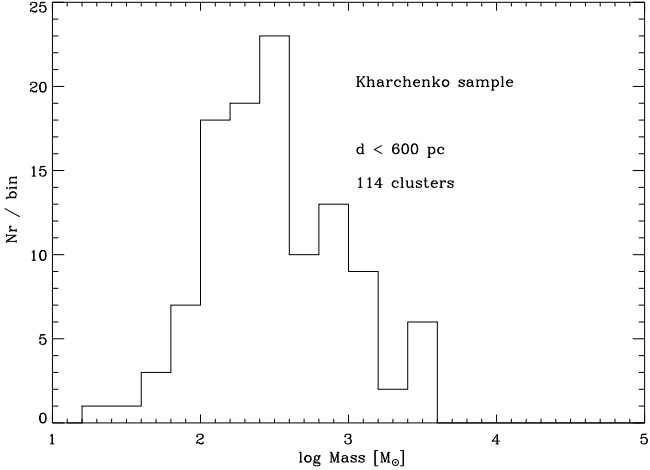,width=9.4cm}}  
\caption[]{
The mass histogram of 114 clusters of the Kharchenko et
al. (2005) catalogue within a distance of 600 pc. The steep edge at
the low mass side suggests that the sample is complete for clusters
with a mass $M \simgreat 100 \Msun$.}
\label{fig:mass_hist}
\end{figure}

The resulting mass-age histogram of the clusters, 
for $M^*_{\rm min}=0.15 \Msun$, is shown in Fig.
\ref{fig:mass_age_hist}. Most of the clusters have a present mass 
in the range of about $5~10^1$ to $5~10^3$ \Msun. 
The histogram of the resulting masses, Fig. \ref{fig:mass_hist}, 
shows a peak in the range of about 100 to 300\Msun. The slow decline
to the high mass end reflects the initial cluster mass function
modified by mass loss and disruption. The steep decrease to low masses
is due to the detection limit of the clusters and their members.
The edge suggests that the mean lower mass limit of the 
Kharchenko et al. (2005) cluster sample is about $100 \Msun$. 
This mass corresponds to a 
minimum number of about 280 stars per cluster if $M^*_{\rm min}=0.15
\Msun$. From Eq. \ref{eq:miapprox} we find that
a present mass of 100 \Msun\ corresponds to an initial mass of
$3.4~10^2$  \Msun\ if $t=10^8$ yr and $5.8~10^3$ \Msun\ if $t=10^9$ yr. 
These values are for a disruption parameter of \tgal= 3.3 Myr (see below).

To estimate the sensitivity of the cluster masses to the
adopted {\it stellar} lower mass limit we also determined the 
masses of the clusters in the Kharchenko et
al. (2005) catalogue within 600 pc in the same way as described above 
but with an adopted minimum stellar mass of $M^*_{\rm min}=0.25 \Msun$
instead of 0.15 \Msun. The resulting cluster masses are about 80\%
of those for $M^*_{\rm min}=0.15\Msun$, so in that case the
minimum lower mass limit of the Kharchenko sample would be about 80 \Msun. 
The limiting cluster mass of 80 \Msun\ corresponds to about 140 stars.


\subsection{The disruption time of clusters in the solar neighbourhood}

\begin{figure}
\centerline{\psfig{figure=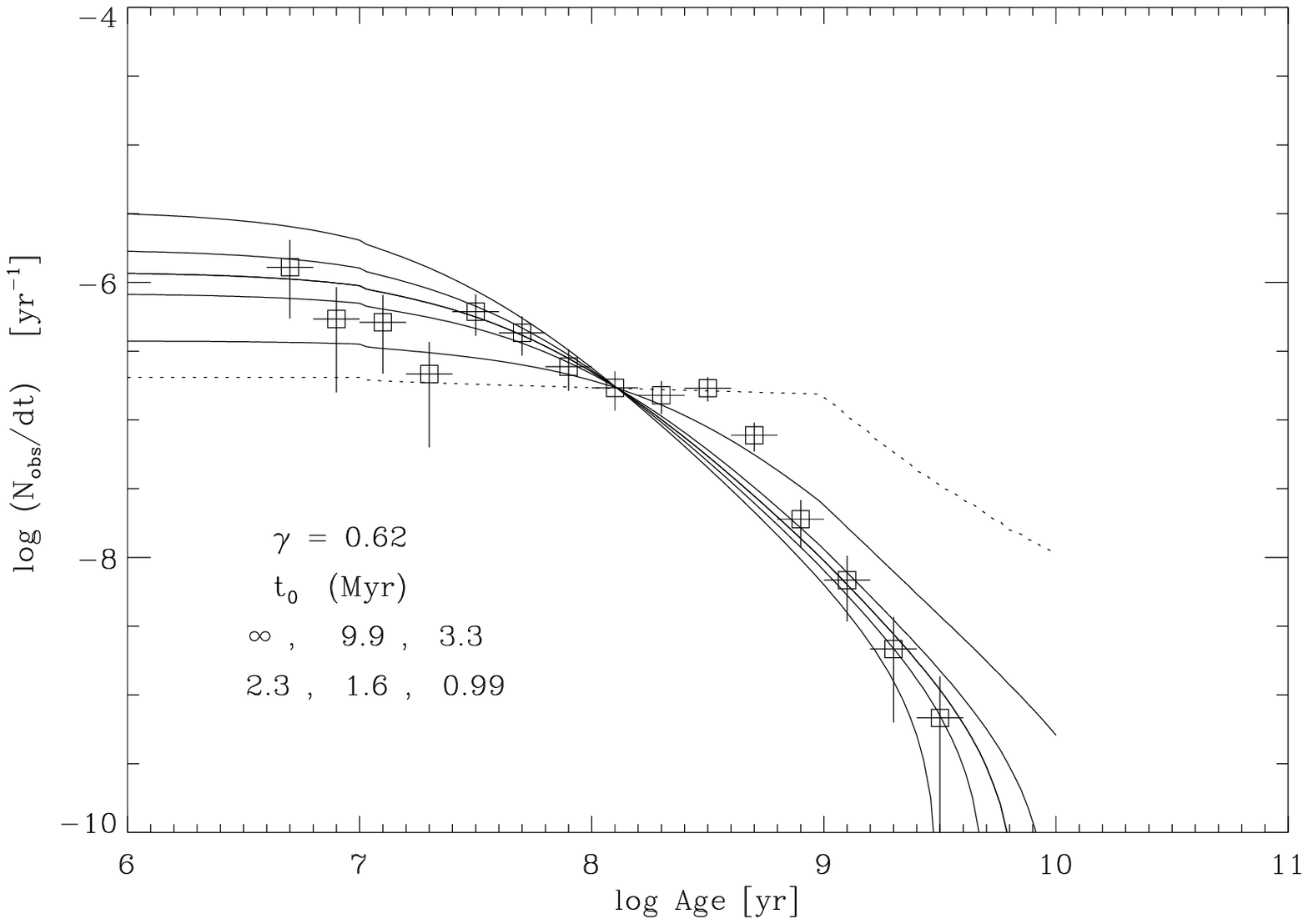,width=9.4cm}}  
\centerline{\psfig{figure=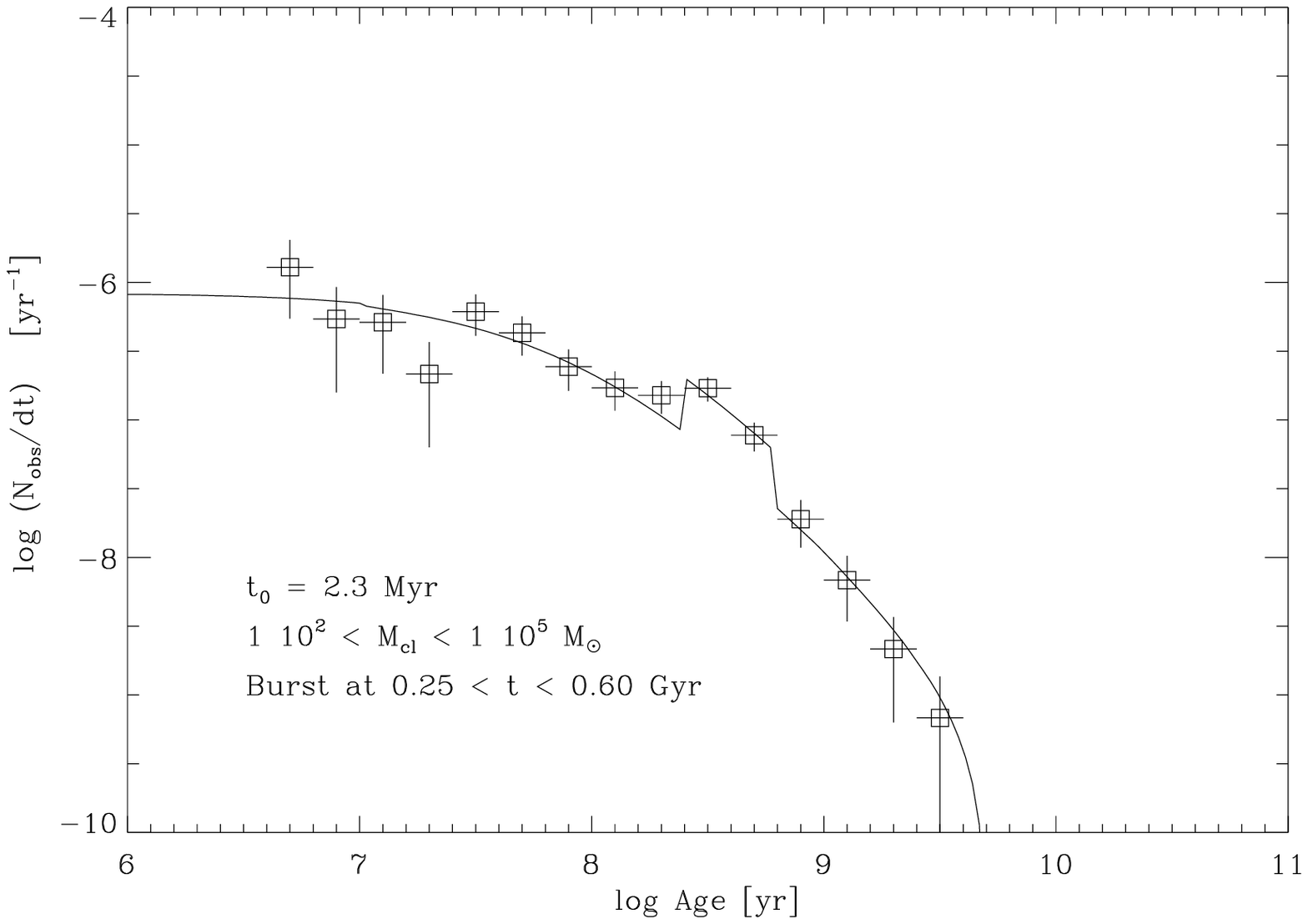,width=9.4cm}}  
\caption[]{ 
Comparison between observed and
predicted  age histogram of clusters in the solar neighbourhood
within $d<600$ pc. The data are fitted to predicted 
relations based on our analytical expression of the cluster
disruption with various values of $\tgal$, 
normalized to the point at
$\log(t/{\rm yr})=8.1$. 
The clusters are formed in the mass range of $10^2 < \Mcl < 10^5$
\Msun, with a CIMF of slope -2.0.
Top figure: Predictions for a constant CFR.
The dotted line indicates the prediction if there was no cluster
disruption, but only mass loss by stellar evolution. 
The shorter the disruption time, the steeper the decrease towards high ages.
At young ages the shortest disruption time corresponds to the largest
formation rate and vice versa.
Lower figure: the best fit for an assumed burst 
between 250 and 600 Myr ago, with a CFR that was 2.5 times higher than
before and after the burst.
}
\label{fig:kharchenkofits}
\end{figure}

Figure \ref{fig:kharchenkofits}  shows the fits to the data 
for $\gamma=0.62$ and various values of $\tgal$, based on the method 
described in Sect. \ref{sec:5.1}. 
For the top figure
we adopted a constant cluster formation rate and a CIMF with
$\alpha=2$ and a mass upper limit of $1~10^5$ \Msun. This latter choice
agrees with the steep decrease around $\log (t/{\rm yr}) \simeq 9.5$.
We assumed a minimum detectable cluster mass of 100\Msun. 
The predicted distributions are
normalized to the data point at $\log (t/{\rm yr})=8.1$, which is one
of the most accurate data points. The best fit is reached for
$\tgal \simeq 3.3$ Myr. (We remind that the low datapoint at
$\log (t/{\rm yr}) =7.3$ and the subsequent high point at 7.5 
may be due to the adopted binning; see Fig. \ref{fig:kharchenko_age})

The distributions for $\tgal >>3.3$ Myr 
underpredict the observed numbers  at $\log (t/{\rm yr}) < 7.2$ 
and overpredicts the numbers at old ages.
The distribution for $\tgal <<1.6$ Myr overpredicts the distribution at
young ages. We have applied a $\chi^2$ test to express the goodness of
the fit. For this test we only considered the data at 
$7.1 \le \log (t/{\rm yr}) \le 8.1$ and $\log (t/{\rm yr}) \ge 8.9$, i.e.
we excluded the bump at $ 8.3 \le \log (t/{\rm yr}) \le 8.7$ which
will be discussed below. We also excluded the data younger than 10
Myr, because they may be affected by infant mortality. The criterion
$\chi^2 \le \chi_{\rm min}^2 +1$ results in a best estimate of
$\tgal=3.1^{+1.2}_{-0.8}$ Myr. 
The data of the fits are given in the top line of Table \ref{tbl:agefits}. 

To investigate the effect of the adopted lower mass limit,
we have also compared the observed age distribution with that
predicted for $M_{\rm max}=1~10^5$ \Msun\ and $M_{\rm min}=80$
\Msun. This last value is derived from the
Kharchenko et al. cluster sample if the stellar lower mass limit of
0.25\Msun\ is adopted instead of 0.15\Msun.
This comparison is very similar to the one in the top
panel of Fig. \ref{fig:kharchenkofits} and is not shown here. 
The resulting data are listed in 
the second line of Table \ref{tbl:agefits}. The best fit is at
a slightly longer disruption time of $\tgal = 3.5^{+1.3}_{-0.8}$ Myr.
Combining the two values of \tgal, derived for $\Mmin$=80 and 100
\Msun, we conclude that $\tgal=3.3^{+1.5}_{-1.0}$ Myr (see bottom line of
Table \ref{tbl:agefits}).

The derived value of \tgal\ implies a total disruption
time of a $10^4$ \Msun\ cluster of  $1.3 \pm 0.5$ Gyr (Eq. \ref{eq:t0t4}).     
This empirically derived disruption time of open clusters is 
about a factor 5 smaller
than the value predicted by the $N$-body simulations of BM03, who predict a
disruption time of 6.3 Gyr for a $10^4$ \Msun\ cluster with an
initial concentration factor of $W_0=5.0$ and 5.9 Gyr if $W_0=7.0$.
The $N$-body simulations of clusters by Portegies Zwart, Hut \& Makino
(1998) resulted in about twice as short disruption times as those of
BM03 (see also Lamers, Gieles \& Portegies Zwart, 2005). This is still
longer than the empirically derived disruption time for open clusters
in the solar neighbourhood.

We consider three possible reasons for this discrepancy:  
(a) the clusters do not start with the initial concentration factors
adopted in the simulations by BM03  and (b) the
presence of another
mechanism (apart from the tidal field) that contributes to the
destruction  of clusters in the solar neighbourhood.

BM03 assumed in their \nbody\ simulations that the
clusters initially fill their tidal radius with a
concentration factor of $W_0=5.0$ or 7.0, defined by King (1966).
These are the values suggested by the 
current density profiles of the globular
clusters. 
However, open clusters are much less centrally condensed  than
globular clusters. Moreover they may not fill their tidal radius when they are
formed. If the clusters are smaller than their tidal radius the internal
relaxation will be faster than predicted and so the disruption
might be faster than predicted. 
 We suggest that open clusters are formed so far out of
equilibrium that they lose a substantial fraction of their mass
within a few crossing times. Most clusters will then disperse
completely, in agreement with the high infant mortality rate. The
surviving clusters might then dissolve along the
lines predicted by the \nbody\ simulations, but on a faster time scale.
$N$-body simulations of clusters with various initial concentration
factors and various initial radii are needed to test this suggestion.

The disruption times calculated by BM03 is an upper limit because the
the values are calculated for tidal disruption in a smooth
 tidal field without other destruction mechanisms.
The destruction of open clusters by encounters with giant
molecular clouds (GMCs) has been proposed by several authors, e.g. Oort
(1958) and  Terlevich (1987).
The problem with this explanation is that disruption by a GMC 
is expected to result in a mass dependence disruption 
of $\tdis \sim M^{\gamma}$ 
with $\gamma=1.0$ (e.g. Spitzer 1987), whereas the observed age distribution
agrees perfectly with $\gamma=0.62$ predicted for tidal effects.
On the other hand, massive clusters are 
likely to have larger tidal radii and hence more interactions 
than low mass clusters and will therefore 
be more susceptible to the influence of GMCs. This might soften the
mass dependence of the disruption time to $\gamma<1$.\\


\begin{table}
\caption[]{The disruption time and the formation rate of clusters 
within 600 pc from the Sun for two assumed values of the lower mass
limit of clusters in the Kharchenko et al. (2005) sample.}
\begin{tabular}{cccc}
\hline\\
Mass range & $\tgal$ & $\log(CFR)$ & $\log(CFR)$ \\
(\Msun)    &  (Myr)  & Nr/yr      & \Msun/yr \\
\hline\\
$8~10^1 - 1~10^5$ & $3.5^{+1.3}_{-1.0}$ &$-6.03\pm 0.02$ & $-3.27\pm 0.02$ \\
 & & & \\
$1~10^2 - 1~10^5$ & $3.1^{+1.2}_{-0.8}$ &$-6.03\pm 0.02$ & $-3.19\pm 0.02$ \\
\hline\\
      adopted     & $3.3^{+1.4}_{-1.0}$ &$-6.03\pm 0.02$ & $-3.23\pm 0.06$ \\
\hline\\
\end{tabular}
\label{tbl:agefits}
\end{table}  


\subsection{The cluster formation rate, the star formation rate and
  the infant mortality rate}

The vertical shift of the predicted relative to the observed age
distribution in the top panel of Fig. \ref{fig:kharchenkofits}
gives the CFR. 
We find a CFR of $0.93 \pm 0.04$ clusters per Myr. (This is about
twice as high as the value derived by Battinelli \& Capuzzo-Dolcetta
(1991) based on the Lyng\aa (1987) catalogue of open clusters brighter
than $M_v=-4.5$.) Our value of the CFR corresponds
to a starformation rate in clusters of
$5.9 \pm 0.8 ~ 10^2$ \Msun/Myr within a region of 600 pc from the
Sun for an CIMF with a slope of $\alpha=2$ and an adopted
lower cluster mass limit of $100 \Msun$.  This
corresponds to a surface formation rate of the galactic disk near the
Sun of $5.2 \pm 0.7~10^{-10}$ \Msun yr$^{-1}$pc$^{-2}$.

This value can be compared with the present total star formation rate in 
the galactic disk near the Sun.
Lada \& Lada (2003) derived a SFR of $7$ to 
$10 \times 10^{-10} \Msun$ yr$^{-1}$pc$^{-2}$ from
embedded clusters in the solar neighbourhood. This value is a factor
1.3 to 1.9 higher than the value derived from the clusters in the Kharchenko
sample. The difference is most likely due to the fact that many of the
star clusters formed in embedded clouds will be dispersed within
10 Myr. At later ages they would not be recognized as clusters.
So if the sample of embedded stars studied by Lada \& Lada (2003) is
complete, the infant mortality rate of clusters in the solar neighbourhood
is about 40 percent. 

We can also estimate the infant mortality rate of the clusters from
the data in the Kharchenko et al. (2005) catalogue. The mean cluster formation
rate in the agebin of $6.6 < \log(t) < 6.8$ is $1.3~10^{-6}$
clusters/yr, whereas it has dropped to $5.4~10^{-7}$ clusters/yr at
$6.8 < \log(t) < 7.0$. This indicates a survival rate of about 40\% and an 
infant mortality rate of about 60\%. The same rates are found if we
compare the mean value of the cluster formation rates in the age bin
of $6.5 < \log(t) < 6.9$ with that of  $6.9 < \log(t) < 7.3$.
So, the comparison of the cluster formation 
rate with the star formation rate of 
Lada \& Lada (2003) and the comparison between the formation
rates of the youngest to the slightly older clusters both suggest
an infant mortality rate of about 50\%.

\subsection{The burst between 250 and 600 Myr ago}

The predictions shown in the top panel of Fig. \ref{fig:kharchenkofits}
and discussed above
 do not explain the bump around $\log (t/{\rm yr}) \simeq
8.5$ which is higher than any of the distributions for a constant
cluster formation rate. So the cluster sample of
Kharchenko et al. (2005) suggests that there was an increased cluster
formation rate around that time. We have modelled this with
Eq. \ref{eq:dnmpt} for several non-constant CFRs. The best fit is shown
in the lower part of Fig. \ref{fig:kharchenkofits}, which was
calculated for $\tgal = 3.3$ Myr and with a CFR that is increased by
0.40 dex between 250 and 600 Myr ago. The fit matches the data
well. This suggests that the cluster formation rate was a factor
2.5 higher during this age range.
Taking into account this burst we find that the mean CFR within 600 pc
from the Sun during the last Gyr in the solar neighbourhood was 
$9.1 \pm 3.5~10^2$ \Msun/Myr which corresponds to a surface formation
rate of $8.1 \pm 3.0~10^{-10}$ \Msun yr$^{-1}$pc$^{-2}$.

Zaritzky \& Harris (2005) found a peak near 400 Myr in the SFR of the SMC. 
During this peak the SFR was at least twice as high as the quiescent
SFR of the SMC. They attribute this peak to the perigalactic passage
of the SMC and the Galaxy. Possibly this passage also triggered
the increased SFR in the galactic disk.

De la Fuente Marcos \& de la Fuente Marcos (2004) have studied the
starformation history in the solar neighbourhood, based on
various open cluster catalogues. They identified five bursts at
0.35, 0.70, 1.13, 1.50 and 1.93 Gyrs respectively. The burst
that we found between 0.25 and 0.6 Gyr may correspond to the one of
0.35 or, more likely,  to a combination of those at 0.35 and 0.70 Gyr found by
de la Fuente Marcos \& de la Fuente Marcos (2004). Our analysis 
of the Kharchenko et al. (2005) cluster sample does
not confirm the other bursts.

\subsection{The upper mass limit of the clusters}

The theoretical fits of the predicted age distribution to the
observed one in Fig. \ref{fig:kharchenkofits} shows that 
the best fit is reached if the 
maximum mass of the clusters formed within 600 pc from the Sun during
the last few Gyr was about $10^5$ \Msun. This upper limit nicely explains the
steep drop in the age distribution at $t \ge 1$ Gyr. 
To check the robustness of this conclusion we have also
calculated models with higher mass upper limits for the clusters.
For instance, if the upper mass limit was $10^6$ \Msun\ then 
we would expect the following predicted and observed numbers of  
clusters in the oldest
logarithmic agebins, indicated in a vector ($\log t_{\rm min},\log
t_{\rm up}$; nr predicted, nr observed): (8.9,9.1;5.1,6), (9.1,9.3;
4.1,3), (9.2,9.4; 3.3,2), (9.3,9.5;2.9,1). We see that in these
oldest agebins the number of observed clusters is within the
statistical $1\sigma$ uncertainty of the number of predicted clusters.
This means that we cannot exclude the possibility that the maximum
cluster mass was higher than $10^6$ \Msun\ and possibly 
as high as $10^6$ \Msun.

An alternative way to consider this point is to find the clusters in
the Kharchenko sample with the highest initial mass. We use the
estimate of the present cluster masses, derived from the number of 
member stars, as described in Sect. 6.1, and than applied
Eq. \ref{eq:miapprox} to convert the present mass into the initial
mass with a disruption parameter $\tgal=3.3$ Myr.
We find that the two initially most massive clusters within 
a distance of 600 pc, 
i.e. the ones with $\{\log(t),\log(M)\}$ = \{9.2,3.5\} and
\{9.4,2.8\} in Fig. \ref{fig:mass_age_hist}, 
 had an initial mass of $2.5~10^4$ and $3.2~10^4$ \Msun\
respectively. For a CIMF with $\alpha=2$ the 
number of cluster decreases with mass as $N\sim M^{-2}$, so the 
expected number of clusters initially more massive than e.g. $10^5$ \Msun\   
will be smaller than 1.

The observed mass upper limit is probably  determined by statistical
effects (Hunter et al. 2003; Gieles et al. 2005b). Its expected value can be 
estimated from the number of observed clusters within 600 pc. For a
CIMF with a slope of -2 the maximum mass expected in the sample is
roughly $M_{\rm max} \simeq N \times M_{\rm in}$, where $N=114$ is the
number of observed clusters and $M_{\rm min}\simeq 10^2 \Msun$. So we
expect a maximum initial mass, set by statistical effects, 
of about $1~10^4$ \Msun. (In reality
it should be slightly higher because the observed number of clusters 
is already affected by disruption.) This agrees quite well with the
maximum initial mass of about $3~10^4$ \Msun\ derived in the previous
paragraph.


\section{Discussion and summary}

We have derived a simple analytical expression for the mass loss from
star clusters due to stellar evolution and disruption as a function of time,
Eq. \ref{eq:muapprox}. This expression agrees excellently with results of
\nbody\ simulations of clusters in the tidal field of our galaxy.
The expression is derived for mass loss by stellar evolution using the
$GALEV$ cluster evolution models (Schulz et al. 2002; Anders \&
Fritze-v. Alvensleben 2003) but can easily be applied to other cluster
evolution models, provided that the mass loss due to stellar evolution
can be expressed by an analytic approximation  (e.g. of the type
proposed in Sect. 3.1). Our analytical expression for the mass loss
from star clusters is different from the one by Vesperini \& Heggie
(1997) because they assumed that tidal effects decrease the mass
of a cluster linearly with time.

Our method is based on the fact that the disruption time of clusters,
defined as $\tdis \equiv (d \ln \Mp / dt)^{-1}$ depends on the mass
\Mp, as $\tdis=\tgal (M/\Msun)^{\gamma}$ with $\gamma=0.62$ for
disruption by two body relaxation in a tidal field. 
This dependence was found
both empirically from a study of cluster samples in four galaxies by
BL03  and by Lamers, Gieles \& Portegies Zwart (2005), and
theoretically from \nbody-simulations by Baumgardt (2001) and BM03.
The description contains a normalization parameter, 
$\tgal$, that depends on the
environment in the parent galaxy of the clusters.
This parameter can vary by more than an order of magnitude between
cluster samples of different galaxies, as shown by BL03.  Simple theoretical
predictions, numerical simulations and
empirical determinations of cluster disruption in a few galaxies
showed that 
$\tgal \simeq C_{\rm env} 10^{-4 \gamma} (\rho_{\rm  amb}/\Msunpc)^{-0.5}$, 
where $\rho_{\rm amb}$ is the ambient density in the galaxy at the 
location of the clusters, and $C_{\rm env} \simeq$ 300 - 800 Myr
(Lamers, Gieles \& Portegies Zwart, 2005). 
The value of \tgal\ may be shorter in interacting galaxies, as
suggested by the study of the star cluster sample of M51 by Gieles et
al. (2005a).

Using our description of mass loss from clusters by stellar
evolution and disruption, we can predict the 
resulting present day mass, \Mp, and age distributions in terms of
$N(\Mp,t)d\Mp dt$ of the surviving
clusters for different cluster initial mass functions and different
cluster formation histories. This is given by Eq. \ref{eq:dnmpt}.
An integration of $N(\Mp,t)$ over mass for different ages gives the
expected age distribution of the surviving clusters. An integration
of $N(\Mp,t)$ over age for different masses gives the expected mass
distribution. This method can be applied to any cluster formation
history!
 
A comparison between predicted and observed distributions 
of selected cluster samples can be used to derive the basic properties
of that cluster population, such as the cluster formation rate,
the cluster IMF, and the disruption parameter $\tgal$.
To demonstrate this method we have applied it to the
age distribution of  open clusters in the solar neighbourhood
within 600 pc based on the new cluster catalogue by
Kharchenko  et al. (2005). Tests showed that the cluster sample in
this catalogue is unbiased up to a distance of at least 600 pc
and possibly 1 kpc. The predicted age distribution  
agrees very well with the empirical one
(Fig. \ref{fig:kharchenkofits}). The main uncertainty in the derived
disruption time $\tgal$ is the unknown lower mass detection limit
of this cluster sample. We estimated this mass limit from the given
number of cluster members brighter than $V=11.5$ magn. and found it to
be between $80$ and $100$ \Msun. With these values, the fit of the
predicted age distribution to the observed one shows that the
disruption time $\tgal$ is $3.3^{+1.4}_{-1.0}$ Myr, which corresponds to
a disruption time of a $10^4$ \Msun\ cluster of $1.3 \pm 0.5$ Gyr. This is
about a factor 5 shorter than predicted by $N$-body simulations of
clusters in the tidal field of the solar neighbourhood (BM03). The difference
is possibly due to the fact that BM03 adopted a rather high initial central
concentration of the clusters that is more applicable to globular
clusters than to open clusters. Moreover encounters with giant
molecular clouds may also shorten the lifetime of open clusters. 

The present star formation rate in the solar neighbourhood within 600
pc, derived in this
paper, is $5.9 \pm 0.8~10^2$ \Msun/Myr during the last few Gyrs. 
This corresponds to a surface formation rate of $5.2\pm 0.7~10^{-10}$
\Msun~yr$^{-1}$pc$^{-2}$, which is about a factor 0.5 to 0.7 smaller than
derived from the formation rate of stars in embedded clusters
(Lada \& Lada, 2003). This suggests that a considerable fraction of the
 embedded clusters will be dispersed before reaching an age of several Myr.

The observed age distribution clearly shows evidence for a bump in the
cluster formation rate between 0.25 and 6 Gyr, when the formation rate
was a factor 2.5 higher than before and later. 
This corresponds to two
of the bumps in the star formation rate derived from cluster samples
by de la Fuente Marcos \& de la Fuente Marcos (2005).
The observed bump  might be due to an
encounter between the SMC  and the Galaxy (e.g. Zaritsky \& Harris, 2005).

Although the upper mass limit of the observed clusters in the solar
neighbourhood is less than about $10^4$ \Msun, we show that this
does not exclude the possible formation of higher mass clusters. Given
a cluster IMF with a slope of $-2$, the absence of clusters more
massive than $10^4$ \Msun\ is in agreement with the statistical
uncertainty, even if the real upper limit for the initial mass 
was as high as $10^6$ \Msun.



\section*{Acknowledgements}

We thank R.-D. Scholz for his collaboration in the production of the  
new open cluster catalogue that was used in this study.
Paul Hodge and Anil Seth are acknowledged for useful discussions and 
suggestions.
We thank the kind referee, Douglas Heggie, for useful suggestions and
critical comments that resulted in an improvement of this paper.
This work is supported by a grant from the Netherlands Research
School for Astronomy (NOVA) to HJGLML and by a grant from the
University of Washington in Seattle where a major part 
of this research was carried out when the first author was
on sabbatical.
N.V. Kharchenko acknowledges financial support by the
DFG grant 436~RUS~113/757/0-1, and the RFBR grant 03-02-04028.


\end{document}